\documentclass{tcibook}
\usepackage{fancyhea}
\usepackage{work}
\usepackage{amsmath,amssymb}
\usepackage{psfrag}
\usepackage{color}
\usepackage{epsfig}
\usepackage{graphicx}               
\usepackage{url}
\usepackage{hyperref}
\usepackage{epstopdf}
\usepackage{multirow}
\usepackage{xspace}
\usepackage{threeparttable}

\newcommand{\nc}{\newcommand}  

\nc{\beq}{\begin{equation}}  
\nc{\eeq}{\end{equation}}  
\nc{\beqa}{\begin{eqnarray}}  
\nc{\eeqa}{\end{eqnarray}}  
\nc{\bea}{\begin{eqnarray}}  
\nc{\eea}{\end{eqnarray}}  
\nc{\ra}{\rightarrow}  
\nc{\lsim}{\begin{array}{c}\,\sim\vspace{-21pt}\\< \end{array}}  
\nc{\gsim}{\begin{array}{c}\sim\vspace{-21pt}\\> \end{array}}  
\nc{\slsh}{\slash\hspace*{-0.22cm}}

\def\to{\rightarrow}
\def\Re{{\cal R \mskip-4mu \lower.1ex \hbox{\it e}\,}}
\def\Im{{\cal I \mskip-5mu \lower.1ex \hbox{\it m}\,}}
\def\be{\begin{equation}}
\def\ee{\end{equation}}
\def\bea{\begin{eqnarray}}
\def\eea{\end{eqnarray}}
\def\bit{\begin{itemize}}
\def\eit{\end{itemize}}
\nc{\eref}[1]{(\ref{#1})}
\nc{\Eref}[1]{Eq.~(\ref{#1})}

\nc{\vev}[1]{ \left\langle {#1} \right\rangle }
\nc{\bra}[1]{ \langle {#1} | }
\nc{\ket}[1]{ | {#1} \rangle }
\nc{\fb}{\,{\rm fb}^{-1}}
\nc{\ev}{{\rm eV}}
\nc{\kev}{{\rm keV}}
\nc{\Mev}{{\rm MeV}}
\nc{\gev}{{\rm GeV}}
\nc{\tev}{{\rm TeV}}
\nc{\mev}{{\rm MeV}}

\def\ee{e^+e^-}

\def\msb{{\bar{\ssstyle M \kern -1pt S}}}

\begin{document}

\def\bibname{References}
\bibliographystyle{plain}

\raggedbottom
\parindent=0pt
\parskip=8pt
\setlength{\evensidemargin}{0pt}
\setlength{\oddsidemargin}{0pt}
\setlength{\marginparsep}{0.0in}
\setlength{\marginparwidth}{0.0in}
\marginparpush=0pt


\renewcommand{\chapname}{chap:intro_}
\renewcommand{\chapterdir}{.}
\renewcommand{\arraystretch}{1.25}
\addtolength{\arraycolsep}{-3pt}

\thispagestyle{empty}






\newpage

\pagenumbering{roman}





\newpage






\chapter{Baryon Number Violation}
\label{chap:pdk}

\begin{center}\begin{boldmath}


{\bf Conveners: K.S.~Babu$^{29}$, E.~Kearns$^2$}

U. Al-Binni$^{50}$,
S. Banerjee$^{32}$,
D.V. Baxter$^{11}$,
Z. Berezhiani$^{12,42}$,
M. Bergevin$^{43}$,
S. Bhattacharya$^{32}$,\\
S. Brice$^8$,
R. Brock$^{21}$,
T.W. Burgess$^{28}$,
L. Castellanos$^{50}$,
S. Chattopadhyay$^{32}$,
M-C. Chen$^{44}$,
E. Church$^{52}$,\\
C.E. Coppola$^{50}$, 
D.F. Cowen$^{30}$,
R. Cowsik$^{54}$,
J.A. Crabtree$^{28}$,
H. Davoudiasl$^3$,
R. Dermisek$^{11}$,\\
A. Dolgov$^{13,17,27,40}$,
B. Dutta$^{36}$,
G. Dvali$^{24}$, 
P. Ferguson$^{28}$,
P. Fileviez Perez$^{20}$,
T. Gabriel$^{50}$,
A. Gal$^9$,\\
F. Gallmeier$^{28}$,
K.S. Ganezer$^5$,
I. Gogoladze$^{45}$,
E.S. Golubeva$^{16}$,
V.B. Graves$^{28}$,
G. Greene$^{50}$,\\
T. Handler$^{50}$,
B. Hartfiel$^5$,
A. Hawari$^{25}$,
L. Heilbronn$^{50}$,
J. Hill$^5$,
D. Jaffe$^3$,
C. Johnson$^{11}$,
C.K. Jung$^{35}$,\\
Y. Kamyshkov$^{50}$,
B. Kerbikov$^{17}$,
B.Z. Kopeliovich$^{39}$,
V.B. Kopeliovich$^{16}$, 
W. Korsch$^{46}$,
T. Lachenmaier$^{41}$,\\
P. Langacker$^{14}$,
C-Y. Liu$^{11}$,
W.J. Marciano$^3$,
M. Mocko$^{19}$,
R.N. Mohapatra$^{47}$,
N. Mokhov$^8$,
G. Muhrer$^{19}$,\\
P. Mumm$^{23}$,
P. Nath$^{26}$,
Y. Obayashi$^{15}$,
L. Okun$^{17}$,
J.C. Pati$^{33}$,
R. W. Pattie Jr.$^{25,38}$,\\
D.G. Phillips II$^{25,38}$,
C. Quigg$^8$,
J.L. Raaf$^8$,
S. Raby$^{37}$,
E. Ramberg$^8$,
A. Ray$^{53}$,
A. Roy$^{18}$,
A. Ruggles$^{50}$,\\
U. Sarkar$^{31}$, 
A. Saunders$^{19}$,
A. Serebrov$^{34}$,
Q. Shafi$^{45}$,
H. Shimizu$^{22}$,
M. Shiozawa$^{15}$,
R. Shrock$^{35}$,\\
A.K. Sikdar$^{53}$,
W.M. Snow$^{11}$,
A. Soha$^8$,
S. Spanier$^{50}$,
G.C. Stavenga$^8$,
S. Striganov$^8$,
R. Svoboda$^{43}$,\\
Z. Tang$^{11}$,
Z. Tavartkiladze$^{10}$,
L. Townsend$^{50}$,
S. Tulin$^{48}$,
A. Vainshtein$^{49}$,
R. Van Kooten$^{11}$,\\
C.E.M. Wagner$^{7,1}$,
Z. Wang$^{19}$,
B. Wehring$^{25}$,
R.J. Wilson$^6$,
M. Wise$^4$,
M. Yokoyama$^{51}$,
A.R. Young$^{25,38}$.

$^1${Argonne National Laboratory, Argonne, IL 60439, USA}\\
$^2${Boston University, Boston, MA 02215, USA}\\
$^3${Brookhaven National Laboratory, Upton, NY 11973, USA}\\
$^4${California Institute of Technology, Pasadena, CA 91125, USA}\\
$^5${California State University at Dominguez Hills, Carson, CA 90747, USA}\\
$^6${Colorado State University, Fort Collins, CO, USA}\\
$^7${Enrico Fermi Institute, University of Chicago, Chicago, IL 60637, USA}\\
$^8${Fermi National Accelerator Laboratory, Batavia, IL 60510, USA}\\
$^9${Racah Institute of Physics, The Hebrew University, 91904 Jerusalem, Israel}\\
$^{10}${Ilia State University, 0162 Tbilisi, Georgia}\\ 
$^{11}${Indiana University, Bloomington, IN 47405, USA}\\
$^{12}${INFN, Laboratori Nazionali Gran Sasso, 67100 Assergi, L$'$Aquila, Italy}\\
$^{13}${INFN, Sezione di Ferrara, Via Saragat 1, 44122 Ferrara, Italy}\\
$^{14}${Institute for Advanced Study, Einstein Drive, Princeton, NJ 08544, USA}\\
$^{15}${Institute for Cosmic Ray Research, University of Tokyo, Kamioka, Gifu, 506-1205, Japan}\\
$^{16}${Institute for Nuclear Research, Russian Academy of Sciences, 117312 Moscow, Russia}\\
$^{17}${Institute for Theoretical and Experimental Physics, 113259 Moscow, Russia}\\
$^{18}${Inter University Accelerator Centre, New Delhi 110067, India}\\
$^{19}${Los Alamos National Laboratory, Los Alamos, NM 87545, USA}\\
$^{20}${Max Planck Institute for Nuclear Physics, Soupfercheckweg 1, 69117, Heidelberg, Germany}\\
$^{21}${Michigan State University, East Lansing, MI 48824, USA}\\
$^{22}${Nagoya University, Nagoya, Aichi 464-8602, Japan}\\
$^{23}${National Institute of Standards and Technology, Gaithersburg, MD 20899, USA}\\
$^{24}${New York University, New York, NY 10012, USA}\\
$^{25}${North Carolina State University, Raleigh, NC 27695, USA}\\
$^{26}${Northeastern University, Boston, MA 02115, USA}\\
$^{27}${Novosibirsk State University, 630090 Novosibirsk, Russia}\\
$^{28}${Oak Ridge National Laboratory, Oak Ridge, TN 37831, USA}\\
$^{29}${Oklahoma State University, Stillwater, OK 74074, USA}\\
$^{30}${Pennsylvania State University, University Park, PA 16802, USA}\\
$^{31}${Physical Research Laboratory, Ahmedabad 380009, India}\\
$^{32}${Saha Institute of Nuclear Physics, Kolkata 700064, India}\\
$^{33}${SLAC National Accelerator Laboratory, 2575 Sand Hill Road, Menlo Park, CA 94025, USA}\\
$^{34}${St. Petersburg Nuclear Physics Institute, Gatchina, 188300 St. Petersburg, Russia}\\
$^{35}${State University of New York at Stony Brook, Stony Brook, NY 11790, USA}\\
$^{36}${Texas A\&M University, College Station, TX 77843, USA}\\
$^{37}${The Ohio State University, 191 W. Woodruff Ave., Columbus, OH 43210, USA}\\
$^{38}${Triangle Universities Nuclear Laboratory, Durham, NC 27710, USA}\\
$^{39}${Universidad T\'{e}cnica Federico Santa Mar\'{i}a, Valpara\'{i}so, Chile}\\
$^{40}${Universit\`{a} degli Studi di Ferrara, Via Saragat 1, 44122 Ferrara, Italy}\\
$^{41}${Universitat Tubingen, 72076, Tubingen, Germany}\\
$^{42}${Universit\`{a} dell$'$Aquila, Via Vetoio, 67100 Coppito, L$'$Aquila, Italy}\\
$^{43}${University of California at Davis, Davis, CA 95616, USA}\\
$^{44}${University of California at Irvine, Irvine, CA 92697, USA}\\
$^{45}${University of Delaware, Newark, DE 19716, USA}\\
$^{46}${University of Kentucky, Lexington, KY 40506, USA}\\
$^{47}${University of Maryland, College Park, MD 20742, USA}\\
$^{48}${University of Michigan, Ann Arbor, MI 48109, USA}\\
$^{49}${University of Minnesota, Minneapolis, MN 55455, USA}\\
$^{50}${University of Tennessee, Knoxville, TN 37996, USA}\\
$^{51}${University of Tokyo, Tokyo, Japan}\\
$^{52}${Yale University, New Haven, CT 06520, USA}\\
$^{53}${Variable Energy Cyclotron Centre, Kolkata 700064, India}\\
$^{54}${Washington University, St. Louis, MO 63130, USA}



\end{boldmath}\end{center}

\centerline{\bf Abstract}

This report, prepared for the Community Planning Study -- Snowmass 2013 -- summarizes the
theoretical motivations and the experimental efforts to search for baryon number violation,
focussing on nucleon decay and neutron--antineutron oscillations. Present and future nucleon 
decay search experiments using large underground detectors, as well as planned neutron--antineutron
oscillation search experiments with free neutron beams are highlighted.

\section{Overview}
\label{sec:pdk_theory}
\newcommand{\ka}{\kappa}
\newcommand{\mcA}{{\mathcal A}}
\newcommand{\mcD}{{\mathcal D}}
\newcommand{\mcF}{{\mathcal F}}
\newcommand{\mcL}{{\mathcal L}}
\newcommand{\mcN}{{\mathcal N}}
\newcommand{\mcO}{{\mathcal O}}
\newcommand{\mcV}{{\mathcal V}}
\newcommand{\mcW}{{\mathcal W}}
\def\Dbarhat{\hat{\makebox[0pt][c]{\raisebox{0.5pt}[0pt][0pt]{$\not$}}\mcD}}
\def\Fbarhat{\hat{\makebox[0pt][c]{\raisebox{0.5pt}[0pt][0pt]{$\not$}}F}}

\newcommand{\PRD}[3]{Phys. Rev. {\bf D#1}, #2 (#3)}
\newcommand{\PLB}[3]{Phys. Lett. {\bf B#1}, #2 (#3)}
\newcommand{\PRL}[3]{Phys. Rev. Lett. {\bf#1}, #2 (#3)}
\newcommand{\NPB}[3]{Nucl. Phys. {\bf B#1}, #2 (#3)}
\newcommand{\vn}{{\vec{n}}}
\newcommand{\vm}{{\vec{m}}}
\newcommand{\si}{\sigma}
\newcommand{\hmu}{{\hat\mu}}
\newcommand{\hnu}{{\hat\nu}}
\newcommand{\hrho}{{\hat\rho}}
\newcommand{\hh}{{\hat{h}}}
\newcommand{\hg}{{\hat{g}}}
\newcommand{\hk}{{\hat\kappa}}

\newcommand{\ord}[1]{\mathcal{O}{(#1)}}
\newcommand{\mPl}{M_{\rm Pl}}

\def\F{{\bf F}}
\def\A{{\bf A}}
\def\J{{\bf J}}
\def\af{{\bf \alpha}}
\def\beqn{\begin{eqnarray}}
\def\eeqn{\end{eqnarray}}

\def\dspace{\baselineskip = .30in}
\def\beq{\begin{equation}}
\def\eeq{\end{equation}}
\def\bea{\begin{equation}}
\def\eea{\end{equation}}
\def\pl{\partial}
\def\na{\nabla}
\def\al{\alpha}
\def\bt{\beta}
\def\Ga{\Gamma}
\def\ga{\gamma}
\def\de{\delta}
\def\De{\Delta}
\def\da{\dagger}
\def\ka{\kappa}
\def\si{\sigma}
\def\Si{\Sigma}
\def\te{\theta}
\def\La{\Lambda}
\def\lam{\lambda}
\def\Om{\Omega}
\def\om{\omega}
\def\ep{\epsilon}
\def\non{\nonumber}
\def\sq{\sqrt}
\def\sqg{\sqrt{G}}
\def\sp{\supset}
\def\sb{\subset}
\def\l{\left (}
\def\r{\right )}
\def\lq{\left [}
\def\rq{\right ]}
\def\fr{\frac}
\def\la{\label}
\def\hs{\hspace}
\def\vs{\vspace}
\def\inf{\infty}
\def\ran{\rangle}
\def\lan{\langle}
\def\ov{\overline}
\def\tl{\tilde}
\def\tm{\times}
\def\lrar{\leftrightarrow}
\def\orvec{\overrightarrow}


Baryon Number, ${\cal B}$, is observed to be an extremely good symmetry of Nature.
The stability of ordinary matter is attributed to the conservation of baryon number.
The proton and the neutron are assigned ${\cal B} = +1$, while their antiparticles have
${\cal B}=-1$, and the leptons and antileptons all have ${\cal B}=0$.  The proton, being
the lightest of particles carrying a non-zero ${\cal B}$, would then be absolutely
stable if ${\cal B}$ is an exactly conserved quantum number.  Hermann Weyl formulated
the principle of conservation of baryon number in 1929 primarily to explain
the stability of matter \cite{Weyl:1929fm}.  Weyl's suggestion was further elaborated by Stueckelberg
\cite{Stueckelberg} and Wigner \cite{Wigner}
over the course of the next two decades.
The absolute stability of matter, and the exact conservation of ${\cal B}$, however,
have been questioned both on theoretical and experimental grounds.  Unlike
the stability of the electron which is on a firm footing as a result of electric
charge conservation (electron is the lightest electrically charged particle), the stability
of proton is not guaranteed by an analogous  ``fundamental" symmetry.
Electromagnetic gauge invariance which leads to electric charge conservation is
a true local symmetry with an associated gauge boson, the photon, while baryon
number is only a global symmetry with no associated mediator.

If baryon number is only an approximate symmetry which is broken by
small amounts, as many leading theoretical ideas elaborated here suggest, it would have
a profound impact on our understanding of the evolution of the Universe, both
in its early history and its late--time future. Violation of baryon number is an essential ingredient
for the creation of an asymmetry of matter over antimatter in a symmetrical
Universe that emerged from the Big Bang \cite{Sakharov:1967dj}.  This asymmetry is a critical
ingredient in modern cosmology, and is the primary driving force for the formation
of structures such as planets, stars, and galaxies, and for the origin of everything they
support.  Even tiny violations of baryon number symmetry would impact the late--time
future of the Universe profoundly.  ${\cal B}$ violation would imply the ultimate instability of the proton
and the nucleus, which in turn would predict the instability of atoms, molecules, planets and
stars, albeit at a time scale of the order of the lifetime of the proton  \cite{Adams:1996xe}.

There are many theoretical reasons to believe that baryon number is perhaps not an exact symmetry
of Nature.  The Standard Model of particle physics is constructed in such a way that ${\cal B}$
is an accidental symmetry of the Lagrangian.  This is true only at the classical level,
however. Quantum effects associated with the weak interactions violate baryon number
non-perturbatively \cite{'tHooft:1976up,'tHooft:1976fv}.  These violations arise because ${\cal B}$
is anomalous with respect to the weak
interactions.  While such violations are so small as to be unobservable (at zero temperature),
owing to exponential suppression factors associated with tunneling rates between vacuua of differing baryon number,
as a matter of principle these tiny effects imply that the minimal Standard Model already has ${\cal B}$ violation.  It is noteworthy that the same ${\cal B}$--violating interactions at high temperature become unsuppressed,
when tunneling between vacuua may be replaced by thermal fluctuations which allow crossing of the barriers \cite{Klinkhamer:1984di,Kuzmin:1985mm}.
It is these high temperature ${\cal B}$--violation of the Standard Model that enables a primordial
lepton asymmetry generated via leptogenesis \cite{Fukugita:1986hr}-- a popular mechanism for generating matter
asymmetry -- to be converted to baryon asymmetry of the Universe.

Within the framework of the Standard Model itself, one can write down higher dimensional operators suppressed
by inverse powers of a mass scale assumed to be much larger than the $W$ and $Z$ boson masses.
Such non-renomrmalizable operators which are fully compatible with the gauge invariance of the Standard Model
indeed lead to baryon number violation at dimension 6, with a suppression factor of
two inverse powers of a heavy mass scale \cite{Weinberg:1979sa,Wilczek:1979hc,Abbott:1980zj}.  What could be the origin of such non-renormalizable operators?
One possible source is quantum gravity \cite{Zeldovich:1976vq,Hawking:1979hw,Page:1980qm,Ellis:1983qm}.
It is suspected
strongly that quantum gravity will not respect any global symmetry such as baryon number.  ${\cal B}$ violating
dimension 6 operators arising from quantum gravity effects would lead to proton decay, but with a long lifetime
estimated to be of order $10^{44}$ yrs.,  which is well beyond the sensitivity of ongoing and near-future
experiments.  Nevertheless, this is a further illustration of non-exactness of baryon number symmetry.

In Grand Unified Theories \cite{Pati,GG}, GUTs for short, which are well motivated on several grounds, baryon number
is necessarily violated, and the proton must decay, albeit with a long lifetime exceeding $10^{30}$ yrs.
These theories unify the strong, weak and  electromagnetic forces into a single unified force.  Simultaneously
these theories also unify quarks with leptons, and particles with antiparticles.
Most remarkably, the unification of the three gauge couplings
predicted by GUTs \cite{Georgi:1974yf} is found to hold, in the context
of low energy supersymmetry, at an energy scale of about $10^{16}$
GeV.  Particles with masses at such a large energy scale (amounting to
distance scale of order $10^{-30}$ cm) are beyond reach of direct
production by accelerators.  Nevertheless, the idea of grand
unification lends itself to experimental test in proton decay, with
the partial lifetime predicted to lie in the range of $10^{30} -
10^{36}$ yrs. for the dominant decay modes $p \rightarrow
\overline{\nu}K^+$ and $p \rightarrow e^+ \pi^0$.  These predictions
are within reach of ongoing and forthcoming experiments.  While proton
decay is yet to be seen, the grand unification idea has turned out to
be spectacularly successful as regards its other predictions and
postdictions. These include an understanding of electric charge
quantization, the co-existence of quarks and leptons and their quantum
numbers, and a natural explanation of the scale of neutrino mass.
Proton decay now remains as a key missing piece of evidence for grand
unification.

One can in fact argue,
within a class of well-motivated ideas on grand unification, that
proton decay should occur at accessible rates, with a lifetime of
about $10^{35}$ years,
for
protons decaying into $e^+ \pi^0$, and a lifetime of
less than a few $\times 10^{34}$ years  for protons decaying into
$\overline{\nu} K^+$.
The most stringent limits on proton lifetimes now come from
Super-Kamiokande~\cite{Kearns}.  It is a remarkable scientific achievement that
this experiment, along with some of its predecessors, has improved the lifetime limit by many
orders of magnitude compared to the pioneering experiment of Reines, Cowan
and Goldhaber of 1954 \cite{Reines:1954pg}.
  For  the two important decay modes mentioned above, the SuperKamiokande limits are \cite{Kearns}:
\begin{equation}
\tau(p \rightarrow e^+\pi^0) > 1.4 \times 10^{34} \; {\rm yrs}, \;\;\;\;\;\;\;\;\;
 \tau(p \rightarrow \bar \nu K^+  )>  5.9 \times 10^{33} \; {\rm yrs}.
 \label{limit}
 \end{equation}
These well-motivated models   then predict the observation of proton decay
if one can improve the current sensitivity (of
Super-Kamiokande) by a factor of five to ten. This is why an improved
search for proton decay, possible only with a large  underground detector,
is now most pressing.

There is another promising way of testing the violation of baryon number,
in the spontaneous conversion of neutrons into antineutrons \cite{Kuzmin:1970vk,Glashow:1979sg,Mohapatra:1980qe}.  This process -- neutron--antineutron
oscillation -- is analogous
to $K^0-\overline{K}^0$ mixing in the meson sector which violates strangeness by two units.  In neutron--antineutron ($n-\overline{n}$) oscillations
${\cal B}$ is
violated by two units.  Thus these oscillations, if discovered, would test
a different sector of the underlying theory of baryon number violation
compared to proton decay searches
which would be sensitive to the $|\Delta {\cal B}| = 1$ sector.

There are several motivations to carry out $n-\overline{n}$ oscillation searches to the next level of
experimental sensitivity. In the Standard Model when small neutrino masses are accommodated via the seesaw mechanism,
lepton number (${\cal L}$) gets violated by two units.  Since the weak interactions
violate ${\cal B}$ and ${\cal L}$ separately at the quantum level, but conserve the difference ${\cal B} -
{\cal L}$, neutrino Majorana mass induced by the seesaw mechanism
is suggestive of an accompanying $\Delta {\cal B} = 2$ interactions, so that $\Delta({\cal B} - {\cal L}) = 0$
is preserved. Secondly, the interactions that lead to $n-\overline{n}$ oscillations may also be responsible for the
generation of the baryon asymmetry of the Universe.  Unlike the leading proton decay modes
$p \rightarrow e^+ \pi^0$ and $p \rightarrow \overline{\nu} K^+$, which violate ${\cal B}$
and ${\cal L}$ by one unit, but preserve ${\cal B} - {\cal L}$ symmetry, $n-\overline{n}$ oscillations
violate ${\cal B} - {\cal L}$ by two units.  The primordial baryon asymmetry induced by interactions
causing such oscillations would survive non-perturbative weak interaction effects, unlike the ones
responsible for the leading modes of proton decay.  There may thus be an intimate connection between
$n-\overline{n}$ oscillations and the observed baryon asymmetry of the Universe \cite{Babu:2006xc}.  Thirdly, unlike the
$|\Delta {\cal B}| = 1$ operator which results in proton decay, the effective operators responsible for
$n-\overline{n}$ oscillations have dimension 9, and are suppressed by five powers of an inverse mass scale.
Consequently, an observation of $n-\overline{n}$ oscillation in the next round of experiments would suggest
a relatively low scale of new physics, of order $10^3 - 10^5$ GeV.  This would be quite contrary to the
grand desert hypothesis.  It should be noted that well motivated quark--lepton unified theories which implement
the seesaw mechanism for neutrino masses at low energy scales do lead to observable $n-\overline{n}$ oscillation
amplitude \cite{Babu:2006xc,Claudson:1983js,Cline:1990bw,Benakli:1998ur,Chung:2001tp,Dutta:2005af,Babu:2006wz,Babu:2008rq,Mohapatra:2009wp,Gu:2011ff,Babu:2013yca,Arnold:2012sd}.

$n-\overline{n}$ oscillation searches have been performed in the past, both with free neutron beams, and within
nuclear environment in large underground detectors.  The best limit on the characteristic time of oscillation
derived from free neutron beam is $\tau_{n-\overline{n}}> 0.86 \times 10^8$ sec. from the ILL experiment at Grenoble
\cite{BaldoCeolin:1994jz}.
The experimental signature of  antineutron annihilation in a free neutron beam is spectacular enough
that an essentially background free search is possible.  An optimized experimental search
for oscillations using free neutrons from a 1 MW spallation target at Fermilab's Project X can
improve existing limits on the free neutron oscillation probability by {\it 4 orders of magnitude}.  This can be
achieved by fully exploiting new slow neutron source and optics technology developed for materials science in
an experiment delivering a slow neutron beam through a magnetically-shielded vacuum to a thin
annihilation target. A null result at this level, when interpreted as $n-\overline{n}$ transition in nuclear matter,
would represent the most stringent limit on
matter instability with a lifetime exceeding 10$^{35}$ $\rm{yrs}$.
Combined with data from the LHC and other searches for rare processes, a null result could also rule out
a class of models where baryogenesis occurs below the electroweak phase transition.

%

\section{Grand Unification and Proton Decay}

\begin{figure}[h!]
\begin{center}
\includegraphics[width=12cm]{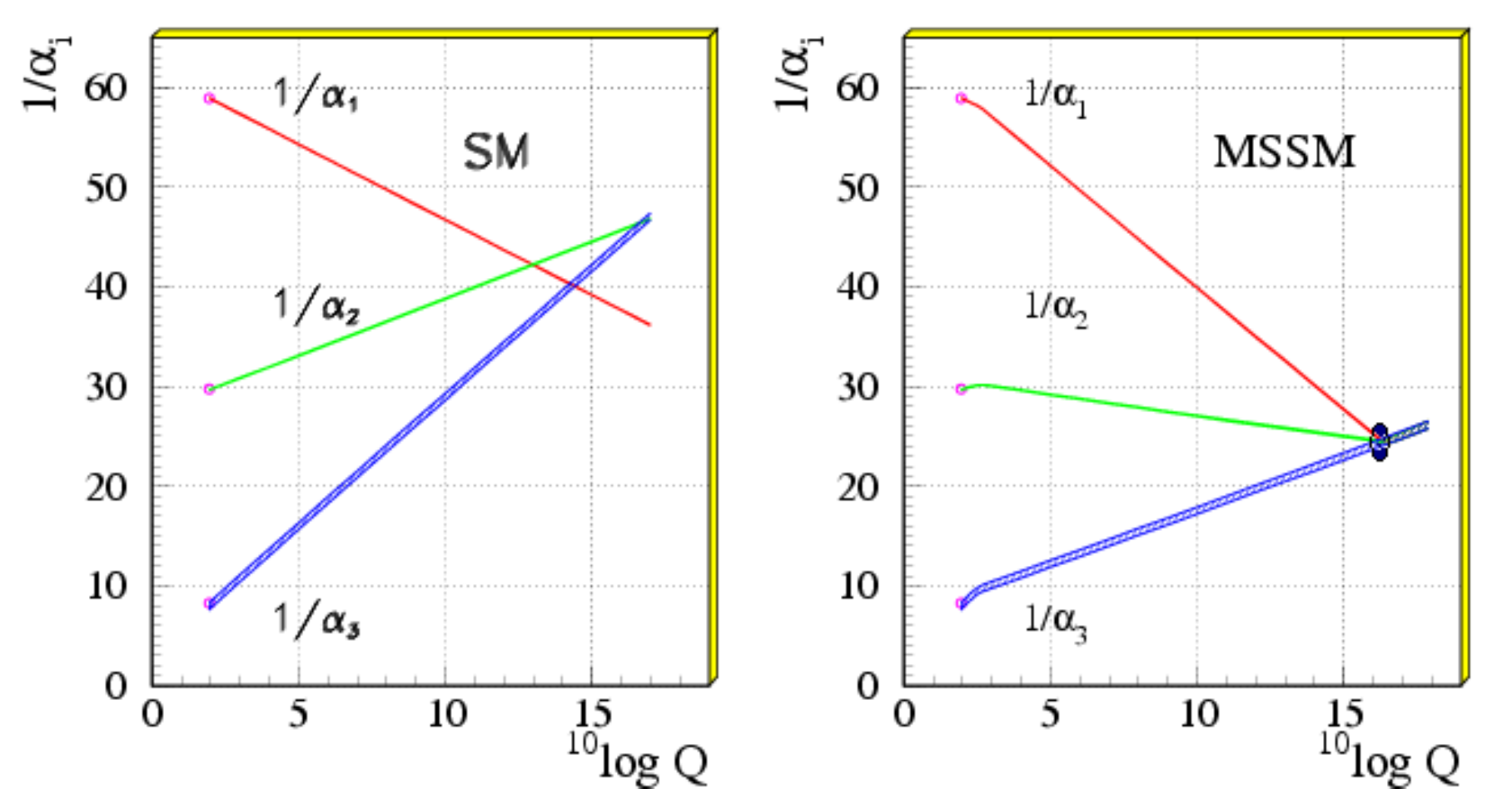}
\caption{\label{unif-diag} Evolution of the three gauge couplings $\alpha_i$ with momentum $Q$:  Standard Model (left panel) and Minimal
Supersymmetric Standard Model (right panel) }\label{unif}
\end{center}
\end{figure}

The most compelling reason for the continued search for proton decay is perhaps the predictions of grand unified
theories \cite{Pati,GG}, which unify the strong, weak and electromagnetic forces into a single underlying force.
The apparent differences in the strengths of these interactions is explained by the running of the coupling
constants with momentum \cite{Georgi:1974yf}.  When the three measured gauge couplings are extrapolated from low energy to higher energies within the context
of the Standard Model particle content, they tend to merge to a common value, but the three do not quite meet
(see Fig. \ref{unif}, left panel.)  With the assumption of low energy supersymmetry, motivated independently
by the naturalness of the Higgs boson mass,
the three gauge couplings are found to unify nicely at a scale $M_{\rm GUT} \approx 2 \times 10^{16}$ GeV (see Fig. \ref{unif}, right panel).
This remarkable feature may be argued as a strong hint in favor of grand unification as well as supersymmetry.

GUTs explain many of the puzzling features observed in Nature that are not explained
by the Standard Model.  A prime example is the quantization of electric charge, together with  $Q_{\rm proton} = -Q_{\rm electron}$
(to better than 1 part in $10^{21}$).  This is a natural consequence of
grand unification, but not of the Standard Model, owing to the non-Abelian nature of the GUT symmetry, which leads to
traceless generators and thus $Q_{\rm proton} + Q_{\rm electron} = 0$.  The co-existence of quarks with leptons
is explained in GUTs which in fact unify these two types of particles.  The miraculous cancellation of chiral
anomalies that occurs among each family of quarks and leptons has a symmetry--based explanation in GUTs.
Furthermore, GUTs provide a natural understanding of the quantum numbers of quarks and leptons.  Certain grand unified theories,
notably those based on the gauge symmetry $SO(10)$ \cite{GeorgiSO(10)}, require the existence of right-handed neutrinos, one per family, which play an essential
role in the seesaw mechanism for generating small Majorana masses for the ordinary neutrinos.  These right-handed neutrinos
also may be responsible for the generation of the baryon asymmetry of the Universe via leptogenesis \cite{Fukugita:1986hr}.
The grouping of quarks with leptons, and particles with antiparticles, in a common GUT multiplet leads to the violation of baryon number
and proton decay \cite{Pati,GG}.  The unification scale of $M_{\rm GUT} \approx 2 \times 10^{16}$ GeV of SUSY GUTs will serve
as the benchmark for proton lifetime estimate.



\subsection{\boldmath{$SU(5)$} Unification and Proton Decay}

$SU(5)$ is the simplest grand unified symmetry that contains the Standard Model gauge symmetry as a subgroup \cite{GG}. It turns out to be the most predictive as regards proton lifetime as well.  This is because of small GUT scale threshold effects arising from the superheavy sector of the theory.  The
masses of these particles, which are the left-over Higgs bosons from the GUT symmetry breaking,
are not precisely determined from the extrapolation of low energy gauge couplings.  Their precise masses affect the determination of the masses
of $X$ and $Y$ gauge bosons of $SU(5)$, which are the mediators of proton decay.  $SU(5)$ being the smallest GUT group has the smallest number
of such supeheavy particles, and thus the least amount of uncertainty in proton lifetime estimate.
The minimal non-supersymmetric version of $SU(5)$~\cite{GG} has already been excluded by the experimental lower limit on $p \rightarrow e^+ \pi^0$ lifetime and the mismatch of the three gauge couplings when extrapolated to high energies
(see left panel of Fig.~\ref{unif-diag}).  With low energy supersymmetry, which is independently motivated by the naturalness of the Higgs boson mass,
there is a simple explanation why the decay $p \rightarrow e^+ \pi^0$ has not been observed.  The unification scale, and hence the mass of
 the $X$ and $Y$ gauge boson that mediate proton decay, increase significantly with low energy SUSY
(see right panel of Fig.~\ref{unif-diag})~\cite{Langacker:1980js}.

\begin{figure}[h!]
\begin{center}
\includegraphics[width=0.75\textwidth]{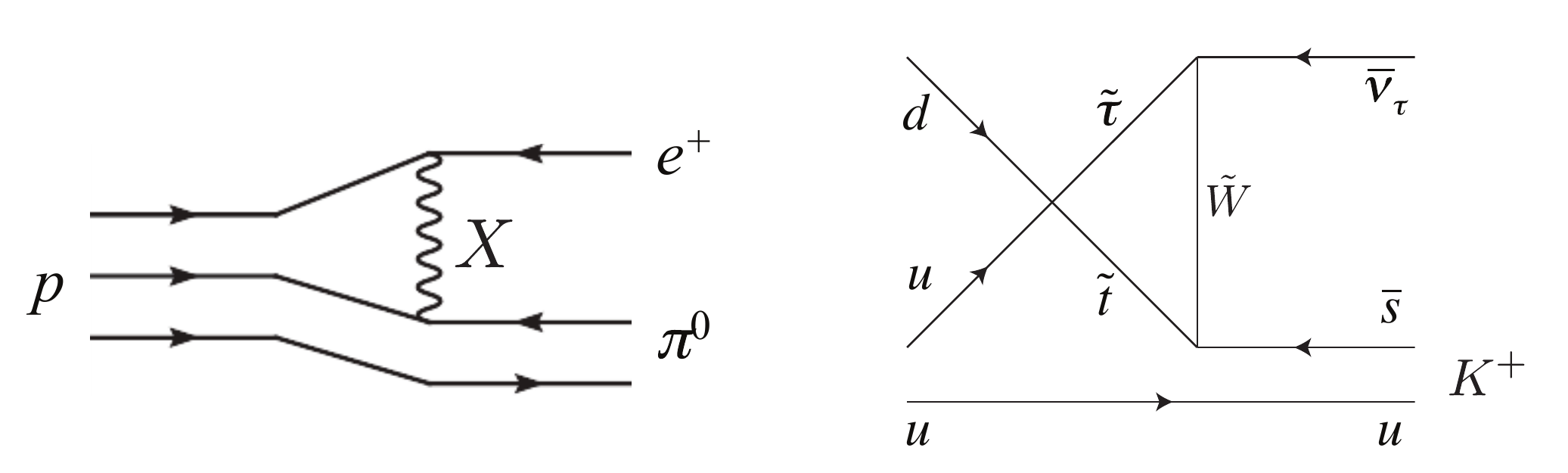}
\caption{\label{pdk-diag} Diagrams inducing proton decay in GUTs.
  $p \rightarrow e^+ \pi^0$ mediated by $X$ gauge boson (left) in non-SUSY and SUSY GUTs, and $p
  \rightarrow \overline{\nu} K^+$ generated by a $d=5$ operator in SUSY GUTs.
  (right).}\label{pdk-diagrams}
\end{center}
\end{figure}

Supersymmetric grand unified theories (SUSY GUTs)~ \cite{Dimopoulos:1981zb,Sakai:1981gr,Dimopoulos:1981dw,sugragut,HLW,Barbieri:1982eh,SUGRA20} are natural extensions of the Standard Model that preserve the attractive features of GUTs noted above, such as quantization of electric charge,  and lead to reasonably precise unification of the three gauge couplings. They also
explain the existence of the  weak scale, which is much smaller than the GUT scale,
and provide a dark matter candidate in the form of the lightest SUSY particle. Low energy SUSY brings in a new twist to proton decay, however, as it predicts a new decay mode $p \rightarrow \overline{\nu} K^+$ that would be mediated by the colored Higgsino~\cite{Sakai:1981pk},\cite{Weinberg:1981wj}, the GUT/SUSY partner of the Higgs doublets (see Fig. \ref{pdk-diag}, right panel).  The lifetime for this mode in minimal renormalizable
SUSY $SU(5)$ is typically shorter than the current experimental lower limit quoted in Eq. (\ref{limit}), provided that the SUSY particle
masses are less than about 3 TeV, so that they are within reach of the LHC.  This is, however, not the case in fully realistic SUSY $SU(5)$ models, as shall be explained below.


In order to evaluate the lifetimes for the $p \rightarrow \overline{\nu}K^+$ and $p \rightarrow e^+ \pi^0$ decay modes in SUSY $SU(5)$~\cite{Hisano:1992jj}, a symmetry breaking sector and a consistent Yukawa coupling sector must be specified.  In $SU(5)$, one family of quarks and leptons is organized as $\{10 + \overline{5} + 1\}$, where $10 \supset \{Q, u^c, e^c\},\overline{5} \supset \{d^c, L\}$, and $1 \sim \nu^c$.  $SU(5)$ contains 24 gauge bosons, 12 of which are the gluons, $W^\pm, Z^0$ and the photon, while the remaining 12 are the $(X,Y)$ bosons that transform as $(3,2,-5/6)$ under $SU(3)_c \times SU(2)_L \times U(1)_Y$ and their conjugates.  These bosons have both diquark and leptoquark couplings, which lead to baryon number violating processes. The diagram leading to the decay $p \rightarrow e^+ \pi^0$ is shown in Fig. \ref{pdk-diag}, left panel.  $SU(5)$ breaks down to the Standard Model symmetry in the supersymmetric limit by employing a ${24_H}$ Higgs boson.  Additionally, a $\{5_H + \overline{5}_H\}$ pair of Higgs bosons is employed, for electroweak symmetry breaking and the generation of quark and lepton masses.  These fields contain the MSSM Higgs fields
$H_u$ and $H_d$.  They also contain color-triplet fields ($T$ and $\overline{T}$) with baryon number violating interactions.  $T$ and $\overline{T}$
must then have a GUT scale mass, while $H_u$ and $H_d$ from the same GUT multiplet must have weak scale masses.  This is done in minimal
SUSY $SU(5)$ by a fine-tuning, which is perhaps an unappealing feature of this theory.

The inverse decay rate for $p \rightarrow e^+ \pi^0$ can be calculated in minimal SUSY $SU(5)$ to be \cite{Hisano:1992jj,SU5,SU5p}
\begin{equation}
\Gamma^{-1}(p \rightarrow e^+ \pi^0) = (1.6 \times
10^{35}~{\rm yr}) \times \left({\alpha_H \over 0.012 ~{\rm GeV}^3}\right)^
{-2}\left({\alpha_G \over 1/25}\right)^{-2}
\left({A_R \over 2.5}\right)^{-2} \left({M_X \over 10^{16}~ {\rm GeV}}\right)^4~.
\end{equation}
Here $M_X$ is the mass of the $X,Y$ gauge bosons that mediate proton decay, $\alpha_G= g_5^2/(4\pi) \simeq 1/25$ where $g_5$ is the unified
gauge coupling, $\alpha_H \simeq 0.01 ~{\rm GeV}^3$ is the nuclear matrix element relevant for proton decay, and $A_R\simeq 2.5$ is the renormalization factor of the effective $d=6$ proton decay
operator. Naively one would expect $M_X$ to be slightly below the unification scale $M_{\rm GUT} \approx 2 \times 10^{16}$ GeV,
since full unification is achieved at momentum scales above the masses of all split multiplets.  Low energy gauge couplings do
provide precise information on a particular combination of GUT scale masses:
\begin{equation}
(-2 \alpha_3^{-1} - 3 \alpha_2^{-1} + 3 \alpha_Y^{-1})(M_Z) =
{1 \over 2 \pi} \left\{36 \,{\rm ln} \left({M_X \over M_Z}
\left({M_\Sigma \over M_X}\right)^{1/3}\right)+ 8 \,{\rm ln}
\left({M_{\rm SUSY} \over M_Z} \right)
\right \}~.
\label{SU(5)-2}
\end{equation}
Here the quantity on the LHS is experimentally measured with high precision.  $M_\Sigma$ on the RHS refers to the mass of
a color octet Higgs field that is left-over from the GUT symmetry breaking.  $M_{\rm SUSY}$ is the effective supersymmetry breaking
mass scale, which is presumed to be known within specific schemes of SUSY breaking.  It is the GUT mass of the color octet
that is uncertain, which reflects in the proton lifetime estimate as well.

The mass of the other super-heavy particle of the theory, the color triplet superfield which mediates
the decay $p \rightarrow \overline{\nu} K^+$, can also be related to low energy observables in analogy
to Eq. (\ref{SU(5)-2}).
In general, agreement with the experimental value of $\alpha_3(M_Z) = 0.1184\pm 0.0007$ demands the color triplet mass to be  lower than
 $M_{\rm GUT} \approx 2 \times 10^{16}$ GeV. This tends to lead to a rate of proton decay into $ \bar{\nu} K^+$ which is in disagreement with observations~\cite{SU5}, at least in the case where the superparticle masses are below about 3 TeV, so that they can be produced at the LHC.

It should be noted, however,  that the Yukawa sector of minimal SUSY $SU(5)$ enters in a crucial way in the rate of proton decay into $\bar{\nu} K^+$. Minimal SUSY $SU(5)$  with renormalizable couplings leads to the relation $M_d = M_\ell^T$, relating the down quark and charged lepton mass matrices.  Consequently, the asymptotic relations $m_b^0 =m_\tau^0,~m_s^0 = m_\mu^0,~m_d^0 = m_e^0$ follow for the masses of quarks and leptons at the GUT scale.  Although the first of these relations agrees reasonably well with observations once it is extrapolated to low energies, the relations involving the two light family fermions are not in agreement with observations.  Allowing for higher dimensional non-renormalizable operators can correct these wrong
relations, however, they will also modify predictions for the decay rate $p \rightarrow \overline{\nu}K^+$ in a way that cannot be precisely
pinned down. Thus the exclusion of the minimal renormalizable SUSY $SU(5)$ model based on the non-observation of the decay $p \rightarrow
\overline{\nu} K^+$ should be taken with some reservation \cite{SU5p}.

Various modifications of the Yukawa sector of minimal SUSY $SU(5)$ that correct the unacceptable fermion mass relations have been studied.
One could introduce new Higgs fields belonging to a $45_H + \overline{45}_H$ representation \cite{Georgi:1979df}, which would however, introduce large GUT
scale threshold corrections.  One could rely on higher dimensional operators to correct the fermion masses, which would also introduce
a large number of parameters. An alternative   possibility, which appears to be simple and predictive, is to add a vector-like pair of $\{5+\overline{5}\}$ fermions~\cite{Babu:2012pb}.  The quarks and leptons from these multiplets can mix differently with the usual quarks and leptons, and thereby correct the bad mass relations $m_s^0 = m_\mu^0$ and $m_d^0 = m_e^0$.  (Such mixings break $SU(5)$ symmetry through the vacuum expectation value of the $24_H$ Higgs field.)
Optimizing these mixings so as to enhance the dominant $p \rightarrow
\overline{\nu} K^+$ lifetime to saturate the present upper limit, approximate upper limits for the various
partial lifetimes are found:
$\tau(p \rightarrow \mu^+ K^0)
\sim 1 \cdot 10^{34}$ yrs, $\tau(p \rightarrow \mu^+ \pi^0)
\sim 2 \cdot 10^{34}$ yrs, and $\tau(p \rightarrow \overline{\nu}+ \pi^0) \sim 7 \cdot 10^{33}$ yrs~\cite{Babu:2012pb}.  In obtaining these numbers, the SUSY particles have been assumed to have masses below 3 TeV, and the unification scale has been taken to be at
least a factor of 50 below the Planck scale, so that quantum gravity effects
remain negligible. Here results of lattice calculations for the nuclear matrix element relevant for proton decay
have been used~\cite{Aoki}.
Since the predicted rates are close
to the present experimental limits, these realistic SUSY $SU(5)$ models can be tested by improving
the current sensitivity for proton lifetime by a factor of ten.

\subsection{\boldmath{$SO(10)$} Unification and Proton Decay}

Models based on $SO(10)$ gauge symmetry are especially attractive since quarks, leptons, anti-quarks, and
anti-leptons of a family are unified  in a single ${\bf 16}$-dimensional spinor representation of the gauge group~\cite{GeorgiSO(10)}.
This explains  the quantum numbers (electric charge, weak charge, color charge)
of fermions, as depicted in Table~1.  $SO(10)$ symmetry contains five independent
internal spins, denoted as $+$ or $-$ signs (for spin--up and spin--down) in Table 1.  Subject to the condition that the number of
down spins must be even, there are 16 combinations for the spin orientations, each corresponding to one fermionic degree. The first three spins denote color charges, while the last two are weak charges.
In addition to the three independent color spins ($r,b,g)$, there is a fourth color (the fourth row), identified as lepton number~\cite{Pati}.
The first and the third columns (and similarly the second and the fourth) are left--right conjugates. Thus $SO(10)$ contains
quark--lepton symmetry as well as parity.  A right--handed neutrino state ($\nu^c$)
is predicted because it is needed to complete the multiplet. Being a singlet of the Standard Model, it naturally acquires a superheavy
Majorana mass and leads in a compelling manner to the generation of light neutrino masses via the seesaw mechanism.
Hypercharge of each fermion follows from the formula $Y = \frac{1}{3} \Sigma (C) - \frac{1}{2} \Sigma (W)$, where  $\Sigma (C)$ is the summation of color spins (first three entries) and $\Sigma (W)$ is the sum of weak spins (last two entries).
This leads to quantization of hypercharge, and thus of electric charge.
Such a simple organization of matter is remarkably beautiful and can be argued as a hint in favor of GUTs based on $SO(10)$.
\begin{table}[h]
{\footnotesize
\begin{center}
\begin{tabular}{||c|c|c|c||}\hline\hline
$u_r:~\{-++~+-\}$ & $d_r:~ \{-++~-+\}$ & $u^c_r:~\{+--~++\}$ & $d^c_r:~ \{+--~--\}$ \\
$u_b:~\{+-+~+-\}$ &  $d_b:~ \{+-+~-+\}$ & $u^c_b:~\{-+-~++\}$ &  $d^c_b:~ \{-+-~--\}$ \\
$u_g:~\{++-~+-\}$ &  $d_g:~\{++-~-+\}$ & $u^c_g:~\{--+~++\}$ &  $d^c_g:~ \{--+~--\}$ \\
$~\nu:~\{---~+-\}$ &  $~e:~~ \{---~-+\}$ & $~\nu^c:~\{+++~++\}$ &  $~e^c:~ \{+++~--\}$ \\ \hline\hline
\end{tabular}
\end{center}
\caption{Quantum numbers of quarks and leptons. The first three
signs refer to color charge, and the last two to weak charge.
To obtain hypercharge, use $Y = \frac{1}{3}\Sigma(C)-\frac{1}{2}\Sigma (W)$.}
}
\end{table}

As in the case of $SU(5)$, when embedded with low energy supersymmetry so that the mass of the Higgs boson is stabilized,
the three gauge couplings of the Standard Model (SM) nearly unify at an energy scale of $M_X \approx 2 \cdot 10^{16}$ GeV in $SO(10)$ models.
The light neutrino masses inferred from neutrino oscillation data ($m_{\nu_3} \sim 0.05$ eV) suggest the Majorana mass of the heaviest of the
three $\nu^c$'s to be $M_{\nu^c} \sim 10^{14}$ GeV, which is close to $M_X$. In a class of $SO(10)$ models discussed further here, $M_{\nu^c} \sim M_X^2/M_{\rm Pl} \sim 10^{14}$ GeV quite naturally.  The lepton number violating decays of $\nu^c$ can elegantly explain the observed baryon asymmetry of the universe via leptogenesis. Furthermore, the unified setup of quarks and leptons in $SO(10)$ serves as a powerful framework in realizing predictive schemes for the masses and mixings of all fermions, including the neutrinos, in association with flavor symmetries in many cases.  All these features make SUSY $SO(10)$ models compelling candidates for the study of proton decay.

Even without supersymmetry, $SO(10)$ models are fully consistent with the unification of the three gauge couplings and the experimental
limit on proton lifetime, unlike non--SUSY $SU(5)$.  This is possible since $SO(10)$ can break to the SM via an intermediate symmetry such as $SU(4)_C \times
SU(2)_L \times SU(2)_R$ \cite{non-SUSYSO(10)}.  Such models would predict that a proton would decay predominantly to $e^+ \pi^0$ with a lifetime in
the range $10^{33} - 10^{36}$ yrs, depending on which intermediate gauge symmetry is realized~\cite{Mohapatra}.  If the intermediate symmetry
is $SU(4)_C \times SU(2)_L \times SU(2)_R \times D$ with $D$ being the discrete parity symmetry, then taking all the relevant threshold effects into
account an upper limit on the lifetime $\tau(p \rightarrow e^+ \pi^0) < 5 \times 10^{35}$ yrs has been derived in Ref. \cite{khan}.  It has been
shown in Ref. \cite{altarelli} that with the intermediate symmetry $SU(4)_C \times SU(2)_L \times SU(2)_R$ (without discrete parity), $SO(10)$ models can also explain the strong CP problem via the axion solution, although $\tau(p \rightarrow e^+ \pi^0)$ in this case can exceed $5 \times 10^{35}$ yrs.

In SUSY $SO(10)$ models, symmetry breaking can occur in two interesting ways.  One type adopts a $\overline{\bf 126}$ of Higgs, a tensor, which couples directly to the $\nu^c$ states and generates large Majorana masses for them.  This class of models has the attractive feature that
the $R$--parity of the Minimal Supersymmetry Standard Model (MSSM), which is
so crucial for identifying the lightest SUSY particle as the dark matter candidate, is an
automatic symmetry, which is part of $SO(10)$.
In this category, a class of minimal $SO(10)$ models employing a single $\overline{\bf 126}$
and a single ${\bf 10}$ of Higgs bosons that couple to the fermions has been developed~\cite{ProtonSO(10)}.
Owing to their minimality,
these models are quite predictive as regards the neutrino  mass spectrum and oscillation angles.  Small quark mixing angles and large neutrino oscillation angles emerge simultaneously in these models, despite their parity at the fundamental level.  The neutrino oscillation angle
$\theta_{13}$ is predicted to be  large in these models.  In fact, this mixing angle was predicted to be $\sin^22\theta_{13} \simeq 0.09$,
well before it was measured to have this central value \cite{BM}.  Proton decay studies of these models~\cite{SO10Pati} show that at least some of the modes among $p \rightarrow \overline{\nu} \pi^+$, $n \rightarrow \overline{\nu} \pi^0$,
$p \rightarrow \mu^+ \pi^0$ and $p \rightarrow \mu^+ K^0$ have inverse decay rates of order  $10^{34}$ yrs,
while that for $p \rightarrow e^+ \pi^0$ is of order $10^{35}$ yrs.  These upper limits are obtained by saturating $\Gamma(p \rightarrow \overline{\nu} K^+)$ with the experimental limit.

The second type of SUSY $SO(10)$ model adopts a set of low-dimensional Higgs fields for symmetry breaking~\cite{Dimopoulos:1981zu}--\cite{Babu:2010ej}.
This includes
spinors ${\bf 16} + {\bf \overline{16}}$, vectors ${\bf 10}$ and an adjoint ${\bf 45}$ which acquires a vacuum expectation value along
the $B-L$ direction of the form $\lan A\ran ={\rm i}\si_2\otimes {\rm Diag}\l a, ~a,~a,~0,~0\r$.  This has quite an interesting effect~\cite{Dimopoulos:1981zu}, \cite{Babu:1994dq},
since it would leave a pair of Higgs doublets from the {\bf 10} naturally light, while giving superheavy mass to the color triplets -- a feature
that is necessary to avoid rapid proton decay -- when the ${\bf 45}$ couples to the vector ${\bf 10}$--plets.
Doublet--triplet splitting is achieved in these models without the need for fine-tuning.
These models predict that
the heaviest of the light neutrinos has a mass that is naturally of order one tenth of an eV, consistent with atmospheric neutrino oscillation data.  This setup
also allows for a predictive system for fermion masses and mixings, in combination with a flavor symmetry.
Models that appear
rather different in the fermion mass matrix sector result in very similar predictions for $p \rightarrow \overline{\nu} K^+$ inverse decay rate, which
has been found to be typically shorter than a few times $10^{34}$ yrs~\cite{Dimopoulos:1981zu}--\cite{Babu:2010ej}.

Recent work in the same class of SUSY $SO(10)$ models which adopt small Higgs representations has shown that there is an interesting correlation between the inverse
decay rates for the $p \rightarrow \overline{\nu} K^+$ and $p \rightarrow e^+ \pi^0$ modes.  The amplitude for the former
scales inversely as the three-halves power of that for the latter, with only a mild dependence on the SUSY spectrum in the
constant of proportionality~\cite{Babu:2010ej}.  This intriguing correlation leads to the most interesting result that the experimental lower limit
of the lifetime for $p \rightarrow \overline{\nu} K^+$ decay provides a theoretical upper limit on the lifetime for $p \rightarrow
e^+ \pi^0$ decay, and vice versa.   An updated version of the work of \cite{Babu:2010ej} has been carried out \cite{BPT13y} by incorporating the latest LHC limits on the masses of the SUSY particles and using a Higgs mass of  126 GeV, while preserving a reasonable degree of SUSY naturalness. As a typical case, one finds:
\begin{eqnarray}
\tau(p \rightarrow e^+ \pi^0) & \leq & 6.8 \times 10^{34} ~{\rm yrs}\,, \nonumber \\
\tau(p \rightarrow \overline{\nu} K^+) & \leq & \left(7 \times 10^{34} ~{\rm yrs}\right) \cdot \left(\frac{\hat{m}_{\tilde{q}}}{2.4~{\rm TeV}}\right)^4
\cdot \left(\frac{550~{\rm GeV}}{m_{\tilde{W}}}\right)^2
\cdot\left(\frac{7}{\tan\beta} \right)^2.
\label{corr}
\end{eqnarray}
Here $\hat{m}_{\tilde{q}}$ stands for a weighted effective mass of the squarks of the three families.
Thus we see that
these predictions are accessible to future experiments, with an improvement in current sensitivity by about a factor of 10.
%

It has been argued that well-motivated supersymmetric GUTs generically predict proton decay rates that can be probed by next-generation experiments.
One could conceive, however, variations of these predictions by either some cancellation of contributions from different ${\cal B}$- and ${\cal L}$-violating dimension-five operators~\cite{ns1}, by  suppression of Higgsino couplings with matter, by judicious choice of the flavor structure~\cite{Dutta:2004zh}, or by the largeness of the scalar masses (see, for example, Ref. \cite{chatto}). For further studies see Ref. \cite{pdkth:Lazarides:1980nt}--\cite{Dermisek:2001hp}, and for a connection between the inflation mechanism and the proton decay rate, see, for example, Ref. \cite{Rehman:2009yj}.

Let us stress in closing that an important prediction of the simple $SU(5)$ and $SO(10)$ GUTs is that proton decay modes obey the selection rule $\Delta ({\cal B}-{\cal L}) =0$ and are mediated by effective operators with dim=6.  
A general effective operator analysis of baryon number violation reveals two conclusions.
One is that  $\Delta {\cal B}\neq 0$ operators with   dim=7 always predict that $|\Delta({\cal B}-{\cal L})| =2$, which leads to decays such as  $n\to e^-+\pi^+$.
These $d=7$ operators that lead to the selection rule $|\Delta({\cal B}-{\cal L})|= 2$ have been shown to arise in $SO(10)$ GUTs \cite{bm12}, and may lead to
observable nucleon decay rates in the non--supersymmetric models with an intermediate scale associated with a symmetry or new particles.
\subsection{Proton Decay in Extra Dimensional GUTs}

 In higher dimensions it is possible to solve the doublet--triplet splitting problem in an elegant way via
 boundary conditions in the extra dimensions~\cite{Kawamura:2000ev}.   In such a setup it is possible to maintain the success of SUSY GUTs as regards the unification  of gauge couplings ~\cite{Hall:2001xr},\cite{Hall:2002ci}.

 String theories 
 that manifest the nice features of 4D SUSY GUTs
 have been constructed with a discrete $Z_4^R$ symmetry.   This symmetry prevents
dimension  3, 4 and 5 lepton and baryon number violating operators, which are potentially dangerous, to all orders in perturbation theory.
The $\mu$-term (the Higgsino mass term in the MSSM) also vanishes perturbatively.  However non-perturbative effects will generate a $\mu$-term of order the SUSY breaking scale, as desired.
 On the other hand,  the low energy theory is guaranteed to be invariant under matter parity.  Thus the lightest SUSY particle is stable and is an
 excellent dark matter candidate.

Nucleon decay in theories with a $Z_4^R$ symmetry~\cite{Gogoladze},\cite{Lee:2010gv} is dominated by dimension 6 operators which lead to the classic decay modes,  $ p \rightarrow e^+ \pi^0, \bar \nu \pi^+$  and $n \rightarrow \bar \nu  \pi^0,  e^+  \pi^-$.   The lifetime for these modes is of order  $\tau \sim \frac{M_C^4}{\alpha_G^2 m_p^5}$  where $M_C$ is the compactification scale of the extra dimension.   This scale is typically less than the 4D GUT scale, {\it i.e.},  $M_C \leq M_{\rm GUT} \approx 2 \times 10^{16}$ GeV.   Thus the rate for nucleon decay in these modes is typically also within the reach of the next generation of experiments. Moreover, due to the absence of $p \rightarrow \overline{\nu} K^+$ decay mode, the observation of proton decay only in the $e^+ \pi^0$ mode may allow one to distinguish minimal four-dimensional unification models from extra-dimensional ones.

\subsection{Induced Nucleon Decay and Asymmetric Dark Matter}

In this subsection we summarize a novel way for nucleon decay, which is motivated by a
desire to understand the dark matter fraction in the Universe, in relation to the baryon
fraction. The energy density of dark matter (DM) is about five times larger than that of
visible baryonic matter (protons and neutrons).
Yet, given that DM has very different properties from baryons,
one may wonder why these two types of matter
have fairly similar shares of the cosmic energy budget.
One explanation is that DM and baryons share a common origin.
The baryon density today originated from an asymmetry between baryons and antibaryons;
when they annihilated, only baryons were left--over.
Similarly, DM may have a primordial asymmetry comparable to the baryon asymmetry in the Universe, so that
when DM particles and antiparticles annihilated during freeze-out, only the residual asymmetric component remained~\cite{ADMRev}.
Typically, this implies that the DM mass is below the weak scale.

If DM and baryons are related, it has been proposed \cite{Davoudiasl:2010am} that
unusual DM signatures involving baryon number violation can arise: DM in the local halo can annihilate nucleons, producing energetic mesons that can be observed in terrestrial experiments.
This process, termed {\it induced nucleon decay} (IND), mimics traditional nucleon decay since the DM states are unobserved.  However, the kinematics of the daughter mesons is different for IND, requiring different analyses compared to existing searches~\cite{Davoudiasl:2011fj}.
In a minimal model, DM is composed of a pair of stable states: a scalar-fermion pair denoted by $\Phi,\Psi$~\cite{Davoudiasl:2010am,Davoudiasl:2011fj} (e.g., these states may be SUSY partners~\cite{Blinov:2012hq}). The DM operator $\Phi \Psi$ may couple to a neutral combination
of light quarks such as $udd$ or $uds$, giving rise to effective interactions between baryons, mesons, and DM.
These interactions lead to inelastic scattering $\Phi \,N \to {\bar \Psi}\, \Pi$ or $\Psi \,N \to \Phi^* \,\Pi$, where $N$ is the nucleon and $\Pi$ is a meson (pion, kaon, etc.).  DM {\it particles} are transmuted into {\it antiparticles}
(${\bar \Psi}, \Phi^*$) by annihilating visible baryons through inelastic scattering.
Since both the initial and final state dark particles are not observed, these processes
resemble nucleon decay into neutrino final states, $N \to \Pi \,\nu$.

For typical values of parameters, the
momentum $p_\Pi$ of the final state meson in an IND scattering is $\sim 0.6-1.4$~GeV.
This provides a kinematic
discriminant between IND and standard nucleon decay (SND), which typically has $p_\Pi \lsim 0.5$~GeV.
The IND scattering rate depends on the hadronic matrix element of the three-quark operator
between a nucleon and a meson.  To get an order-of-magnitude estimate for the effective lifetime
of a nucleon within the local DM halo,
Refs.~\cite{Davoudiasl:2010am,Davoudiasl:2011fj} used chiral perturbation theory and found
\beq
{\tau^{\rm eff}_{N} \approx 10^{32} \; \text{yr} \times \left( \frac{\Lambda_{\rm IND}}{1 \; \text{TeV} }
\right)^{6}\left( \frac{0.3 \; \text{GeV}/\text{cm}^3 }{\rho_{\rm DM}} \right)} \,,
\label{taueffN}
\eeq
where $\Lambda_{\rm IND}$ is the mass scale of the quark-DM nonrenormalizable coupling and $\rho_{\rm DM}$ is the local energy density of DM.
It should be cautioned that chiral perturbation theory is valid for $p_\Pi \ll 1$~GeV and more precise estimates
require fully non-perturbative lattice calculations \cite{lattice}.
Given that new TeV-scale physics may play a role in electroweak symmetry breaking, one could reasonably assume that $\Lambda_{\rm IND} \gsim 1$~TeV.
It is quite interesting that the estimate of Eq. (\ref{taueffN}) for $\tau^{\rm eff}_N$ is close to the current and future limits on nucleon lifetimes $\tau_N$ from
SND experiments.

IND searches have a natural complementarity with LHC searches for DM.  The operator that connects light quarks ($q$) and DM can mediate monojet processes of the type $q \, q \to {\bar q} + E_T^{\rm miss}$, where $E_T^{\rm miss}$ is missing transverse energy from DM escaping the detector.
The estimates in Ref.~\cite{Davoudiasl:2011fj}
suggest that the LHC at $\sqrt{s} = 14$~TeV and with $\ord{100~{\rm fb}^{-1}}$ of integrated luminosity
can be sensitive to $\Lambda_{\rm IND} \sim 1-4$~TeV, albeit with some model dependence resulting from
underlying ultraviolet physics.
Hence, given the estimate of Eq. (\ref{taueffN}),
one would typically expect a direct correlation between observable signals of IND at nucleon decay experiments and at the LHC.





\subsection{Gauging Baryon Number}

Grand Unified Theories, while elegant, must be realized at a very high energy scale of order $10^{16}$ GeV, well beyond the reach of
foreseeable accelerators for direct detection of new particles.  It is possible to understand the approximate conservation of baryon
number by making ${\cal B}$ a {\it local} symmetry ~\cite{Pais:1973mi,Pati:1975ca}.  This would require additional fermions for anomaly cancelation,
since ${\cal B}$ is an anomalous symmetry with respect to the weak interactions.  In such theories with gauged ${\cal B}$, spontaneous
symmetry breaking may occur at relatively low energies, which may be accessible to colliders. Model building with the inclusion of new
fermions in order to gauge ${\cal B}$ has been  studied in great detail in Refs.~\cite{FileviezPerez:2011pt, Duerr:2013dza}.
In these papers it has been shown that simple extensions of the SM with a local ${\cal B}$ symmetry broken at the TeV scale are fully
consistent with experimental constraints, including cosmology. This class of theories could open a new path to probing the origin of
approximate ${\cal B}$ conservation.

\section{Nucleon Decay Experiments: Past, Present and Future}
\label{sec:pdk_exp}
When Grand Unified Theories were formulated in the mid-1970's, they
predicted lifetimes of the nucleon as short as $10^{29}$ years, which put
experimental detection within range of kiloton scale detectors. First
generation experiments were quickly proposed and mounted: IMB and
Soudan in the United States, Kamiokande in Japan, NUSEX and Frejus in
Europe. None of these experiments found a significant signal, and
notably, IMB and Kamiokande excluded the minimal $SU(5)$ prediction for
the decay $p \rightarrow e^+\pi^0$. The limits from these experiments
populate roughly 70 exclusive decay modes tabulated by the Particle
Data Group \cite{PDG-nucleon-decay-modes}.

The second generation is singularly comprised of the Super-Kamiokande
experiment, which began in 1996 and is ongoing. Super-Kamiokande also
has never yet found a significant nucleon decay signal. The experiment
has extended the lifetime limits for numerous modes by more than an
order of magnitude over the first generation experiments. As of early
2013, the Super-Kamiokande collaboration have recorded 260
kiloton$\cdot$years of exposure with partial lifetime limits in the
range $10^{33}$ to $10^{34}$ years in many cases. The results from
Super-Kamiokande provide both a baseline for comparison and challenge
to the next generation experiments. The next subsection reviews the
results from Super-Kamiokande.

\subsection{Super-Kamiokande Searches for Nucleon Decay}
\label{sec:superk-pdk}

The Super-Kamiokande water Cherenkov experiment dominates the current
limits set on the lifetime of the proton and bound neutron. The
22,500-ton fiducial mass of the detector has
$7.5\times10^{33}$~protons and $6.0\times10^{33}$~neutrons. Fully
contained atmospheric neutrino interactions in the GeV range
constitute the background. The experiment has been collecting data
since 1996 with four distinct data-taking periods called SK-I, -II,
-III, and -IV.  During the SK-I, -III, and -IV periods, $\sim$11,000
inward-facing 20-inch photomultiplier tubes (PMTs) were distributed
evenly on the entire inner detector (ID) surface to provide 40\%
photocathode coverage. An accident that destroyed roughly half of the
PMTs in the inner detector led to the the SK-II period, where the
remaining functional PMTs were redistributed evenly across the ID
surface with approximately 20\% photocathode coverage. This SK-II
period of reduced coverage is notable for future generation water
Cherenkov detectors because the Super-K nucleon decay and atmospheric
neutrino analyses show that the reduction in photocathode coverage
does not have a large adverse effect on nucleon decay detection
efficiency or background rejection. Photodetection drives the cost of
large water Cherenkov detectors, and containing this cost is of great
importance. The SK-IV period introduced new electronics, and
incremental gains in detection efficiency or background rejection are
seen in some analyses.

There are several methods of searching for nucleon decay.  The most
straightforward method is to define a set of selection criteria that
maximize the signal detection efficiency and minimize the
background. The $p\rightarrow e^{+}\pi^{0}$ mode is a good example of
this technique. The proton decay signal is simulated using a Monte
Carlo program that includes effects due to Fermi motion, nuclear
binding potential, intranuclear reactions, and correlated nucleon
effects (all of which are absent for the free proton that is the
hydrogen nucleus). Deleterious nuclear effects are largely responsible
for the overall signal efficiency of roughly 40\%. The background is
estimated by a simulated 500 year exposure of atmospheric neutrinos
using the detailed model that is standard for Super-K, one that takes
into account the atmospheric neutrino flux, neutrino-nucleus cross
sections, and interactions within the nucleus. The signal efficiency
and background rates are summarized in
Table~\ref{tab:sk-eppi0-effbg}. The systematic uncertainty in the
efficiency is estimated to be 19\% and the systematic uncertainty in
the background rate is estimated to be 44\%, independent of SK
period. The background rate has been independently checked using
GeV-scale muon neutrino data in the 1-kiloton water Cherenkov detector
that served as a near detector in the K2K long-baseline neutrino
oscillation experiment \cite{Mine:2008pr}. The preliminary Super-K
result for $p \rightarrow e^+\pi^0$, as of 2013, is detection of zero
candidate events and a 90\% C.L. limit on the partial lifetime of
$1.4 \times 10^{34}$ years \cite{Shiozawa-TAUP13}.

\begin{table}[htb]
\caption{Signal efficiency and background rates for the $p \rightarrow
e^+\pi^0$ analyses by Super-Kamiokande. Uncertainties listed are
statistical, due to the Monte Carlo event sample. These are preliminary
numbers based on improved and updated analyses from those published
in 2011 \cite{Nishino:2012ipa}.}
\label{tab:sk-eppi0-effbg}
\begin{center}
\begin{tabular}{r|c|c|c|c} \hline\hline
 & data & $p \rightarrow e^+\pi^0$ & atmos. $\nu$ & atmos. $\nu$ \\
 & livetime & signal efficiency & estimated bkg. & bkg. rate (evts/Mt/y) \\
\hline
SK-I   & 91.7 kt y & $39.2 \pm 0.7\%$ & 0.27 evts. & $2.9 \pm 0.6$ \\
SK-II  & 49.2 kt y & $38.5 \pm 0.7\%$ & 0.15 evts. & $3.0 \pm 0.5$ \\
SK-III & 31.9 kt y & $40.1 \pm 0.7\%$ & 0.07 evts. & $2.3 \pm 0.6$ \\
SK-IV  & 87.3 kt y & $39.5 \pm 0.7\%$ & 0.22 evts. & $2.5 \pm 0.6$ \\
K2K    & & & & $1.63^{+0.42}_{-0.33}$(stat)$^{+0.45}_{-0.51}$(sys)\\
\hline
\hline
\end{tabular}
\end{center}
\end{table}

\begin{figure}[htb]
\begin{center}
\label{fig:superk-eppi0}
\includegraphics[width=1.0\textwidth]
      {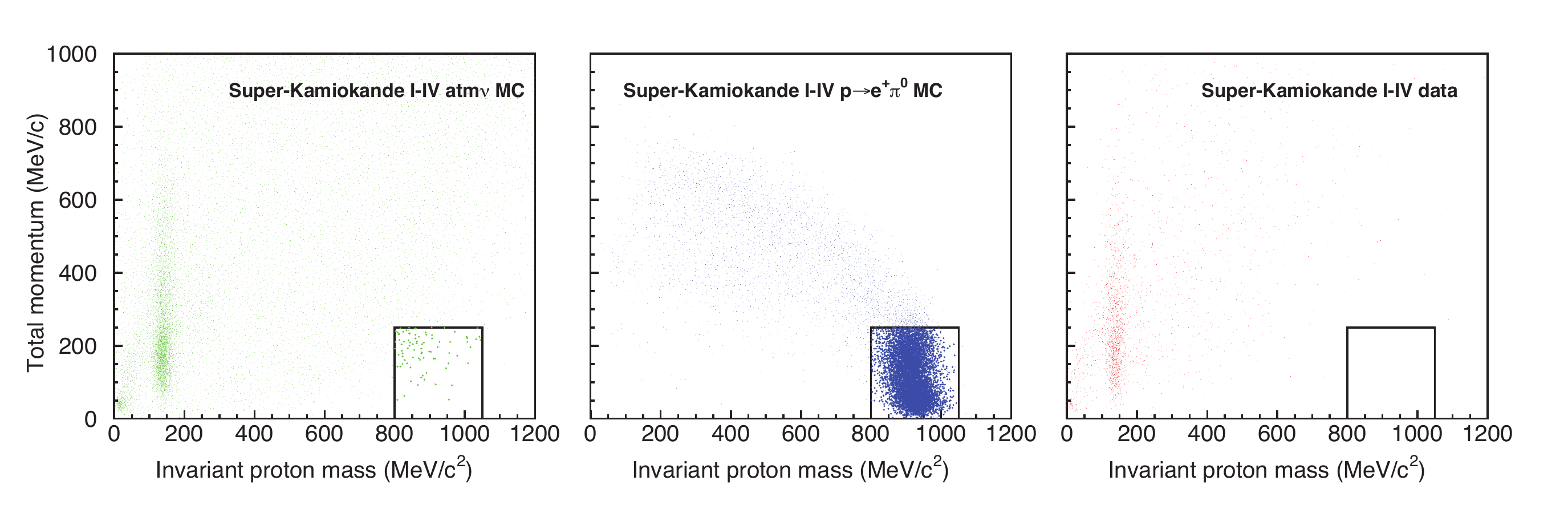}
\caption{The combined
    SK-I+II+III+IV analysis of $p \rightarrow e^+\pi0$.}.
\end{center}
\end{figure}

As can be seen in Fig.~\ref{fig:superk-eppi0}, the signal region is
defined by a box indicating the expected ranges of total reconstructed
momentum and invariant proton mass. The background events that pass
all other selection cuts (atmospheric neutrino interactions) do not
typically fall into the range of momentum and invariant mass that one
expects for proton decay events, making this a low-background search
mode. As long as background rates are kept small (no more than a few
events, preferably less than one), future large water Cherenkov
detectors will extend these limits by a factor of the increase in
detector mass times running time. Also, for searches with background
estimates substantially lower than 1 event, discovery of proton decay
by a single clean event remains possible.

A second technique is used for some decay modes in which a low
background cannot be achieved. For these modes, a ``bump search'' is
done.  Example of this are $n\rightarrow \nu\pi^{0}$ and
$p \rightarrow \nu \pi^+$, where one looks for a momentum peak from
mono-energetic single pions on top of a background consisting mostly
of atmospheric neutrino events with a single $\pi^{0}$ or
non-showering ring\cite{Abe:2013lua}. For this type of search,
understanding the shape of the background event spectrum is
critical. Because these searches are inherently background limited,
future large water Cherenkov detectors will not greatly extend the
sensitivity.

The search for the SUSY GUT favored $p \rightarrow
\overline{\nu}K^{+}$ mode is basically a search for $K^+$ decay at rest,
as the kaon from proton decay is below the Cherenkov threshold. For
the kaon decay to $\pi^+\pi^0$ (branching ratio of 21\%), selection
criteria are used as described above. For the largest branching ratio,
$K^+ \rightarrow \mu^+\nu_\mu$ (64\%), Super-K uses a combination of
the bump search technique plus a low-background cut-based analysis
that tags events by a low energy photon from the de-excitation of the
excited nucleus that is left after the decay of a proton in
$^{16}$O. Using a combination of these techniques allows the
measurement to push the limit on the proton decay lifetime further
than using any of the individual methods. The combined efficiency and
background rates for the low-background selections of $\pi^+\pi^0$ and
nuclear-$\gamma$ tag are shown in Table~\ref{tab:sk-nukp-effbg}. The
weaker performance of SK-II with 20\% photocoverage is now evident,
due to decreased ability to separate 6 MeV nuclear gamma rays from
background. Improved performance due to upgraded electronics in SK-4
is also evident, mainly due to increased efficiency for identifying
Michel electrons.

\begin{table}[htb]
\caption{Signal efficiency and background rates for the $p \rightarrow
\nu K^+$ analyses by Super-Kamiokande. Uncertainties listed are
statistical, due to the Monte Carlo event sample. These are
preliminary numbers based on improved and updated analyses from those
reported as late as 2012. The efficiencies and background rates are
for the combination of the relatively low-background techniques:
$K^+ \rightarrow \pi^+\pi^0$ plus
$K^+ \rightarrow \nu_\mu \mu^+$ with nuclear-$\gamma$ tag.}
\label{tab:sk-nukp-effbg}
\begin{center}
\begin{tabular}{r|c|c|c|c} \hline\hline
 & data & $p \rightarrow \nu K^+$ & atmos. $\nu$ & atmos. $\nu$ \\
 & livetime & signal efficiency & estimated bkg. & bkg. rate (evts/Mt/y) \\
\hline
SK-I   & 91.7 kt y & $15.7 \pm 0.2\%$ & 0.3 evts. & $2.8 \pm 0.4$ \\
SK-II  & 49.2 kt y & $13.0 \pm 0.2\%$ & 0.3 evts. & $6.2 \pm 0.8$ \\
SK-III & 31.9 kt y & $15.6 \pm 0.2\%$ & 0.1 evts. & $3.1 \pm 0.5$ \\
SK-IV  & 87.3 kt y & $19.1 \pm 0.2\%$ & 0.3 evts. & $3.5 \pm 0.4$ \\
\hline
\hline
\end{tabular}
\end{center}
\end{table}

\begin{figure}[htbp]
\begin{center}
  \includegraphics[width=0.8\textwidth]
{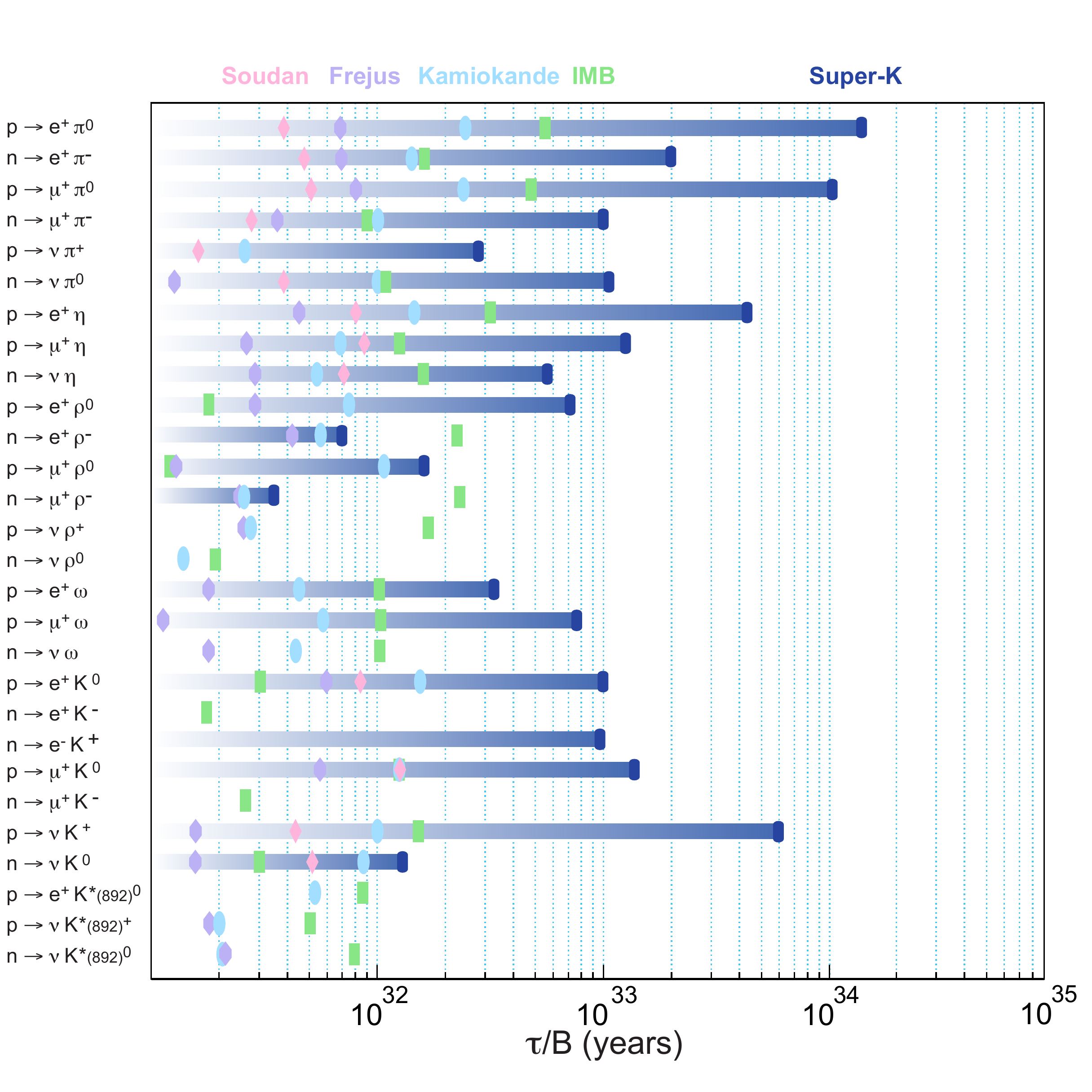}
\end{center}
\caption{\label{fig:superk_pdk}
Summary of lifetime limits for proton or bound neutron
decay into antilepton plus meson; the complete set of possible
two-body decay modes that conserve $B-L$ is listed. Experimental
searches were conducted by Super-K (dark blue gradient band with
marker) and previous experiments: Soudan (pink diamonds), Frejus
(purple hexagons), Kamiokande (light blue ovals), and IMB (light green
rectangles).}
\end{figure}

No signs of nucleon decay have been seen yet in any Super-K analysis
of any nucleon decay mode. The 90\% C.L. lifetime limits of Super-K
nucleon decay searches to antilepton plus meson are summarized in
Fig~\ref{fig:superk_pdk}, compared with measurements from past
experiments. These are two-body decay modes that conserve $(B-L)$. As
mentioned previously, although many GUTs automatically predict one or
more of these modes with varying branching ratios, it is important to
also consider three-body decay modes and alternative origins of baryon
number violation that conserve $(B+L)$\footnote{ Decay modes with a
final state neutrino, always unobserved, may be interpreted as
conserving $(B-L)$ or $(B+L)$} or violate only $B$ (e.g. dinucleon
decay).  Studies along these lines are an active area of inquiry
within the Super-Kamiokande collaboration and a handful of first
results, all negative so far, are presented in talks, theses, or are
being prepared for publication.

\subsection{Proposed Proton Decay Search Experiments}
\label{sec:pdk_proposed_exp}
There are a variety of proposals to continue the search for nucleon
decay with a new generation of experiments. Some of these proposals
are inactive or discontinued, while others are being actively
discussed in various parts of the world. The proposed detectors can be
categorized broadly in three distinctive technologies: water Cherenkov
detectors, liquid argon TPCs and scintillator detectors.
Table~\ref{tab:NNN_detectors} shows a summary of the proposed projects
categorized by technology and region. Not included are some large
detectors that may be able to contribute but that have not developed
the case for proton decay (for example multi-kiloton reactor
experiments using liquid scintillator).

\begin{table}[htb]
\caption{Next generation nucleon decay detector proposals. Inactive
efforts are marked with an asterisk. Speculative future projects that
are less fully-developed and that may be considered for a future
generation are marked with a dagger. 
\label{tab:NNN_detectors}}
\begin{center}
\begin{tabular}{c|c|c|c|c} \hline\hline
Name & Technology & Mass & Location & Reference \\ \hline
Hyper-K & WC & 560 kt & Japan & \cite{Abe:2011ts} \\ \hline
LBNE & LArTPC & 10-70 kt & U.S. & \cite{Akiri:2011dv} \\ \hline
LENA & Scintillator & 50 kton & Europe & \cite{Wurm:2011zn} \\ \hline
GLACIER & LArTPC & 20-100 kt & Europe & \cite{Badertscher:2010sy} \\ \hline
MEMPHYS & WC & 500 kton & Europe & \cite{MEMPHYS} \\ \hline
WbLSc & Scintillator & 23 kton & Japan & \cite{WbLSc} \\ \hline
LBNE-WC & WC & 100-200 kton & U.S. & *\cite{Akiri:2011dv} \\ \hline
UNO & WC & 440 kton & U.S. & *\cite{UNO} \\ \hline
LANDD & LArTPC & $n \times$ 5 kton & U.S. & *\cite{Cline:2006st} \\ \hline
MICA & WC (ice) & multi-Mton & South Pole & $\dagger$ \cite{MICA} \\ \hline
TITAND & WC & multi-Mton & ocean & $\dagger$ \cite{Suzuki:2001rb} \\ \hline
\hline
\end{tabular}
\end{center}
\end{table}

In the following sections, the search for nucleon decay in each of the
three categories: water Cherenkov, liquid argon TPC (LArTPC), and
scintillator are described, with an emphasis on the leading candidate
experiment for each technology.

\subsection{Water Cherenkov}

A next-generation underground water Cherenkov detector,
Hyper-Kamiokande (Hyper-K), is proposed for location in the Kamioka
mine in Japan.  It will serve as the far detector of a long baseline
neutrino oscillation experiment using the off-axis J-PARC neutrino
beam, and as a detector capable of observing nucleon decays,
atmospheric neutrinos, and neutrinos from astronomical origin. The
baseline design of Hyper-K is based on the highly successful Super-K
experiment, taking full advantage of well-proven methods and
technology. The fiducial mass of the detector is planned for 0.56
million metric tons, which is about 25 times larger than that of
Super-K. The details of the proposed experiment can be found in a
published Letter of Intent~\cite{Abe:2011ts}. The Hyper-Kamiokande group has
had several open organizational meetings over the past two years.

\begin{figure}[htbp]
  \begin{center}
    \includegraphics[width=0.85\textwidth]
     {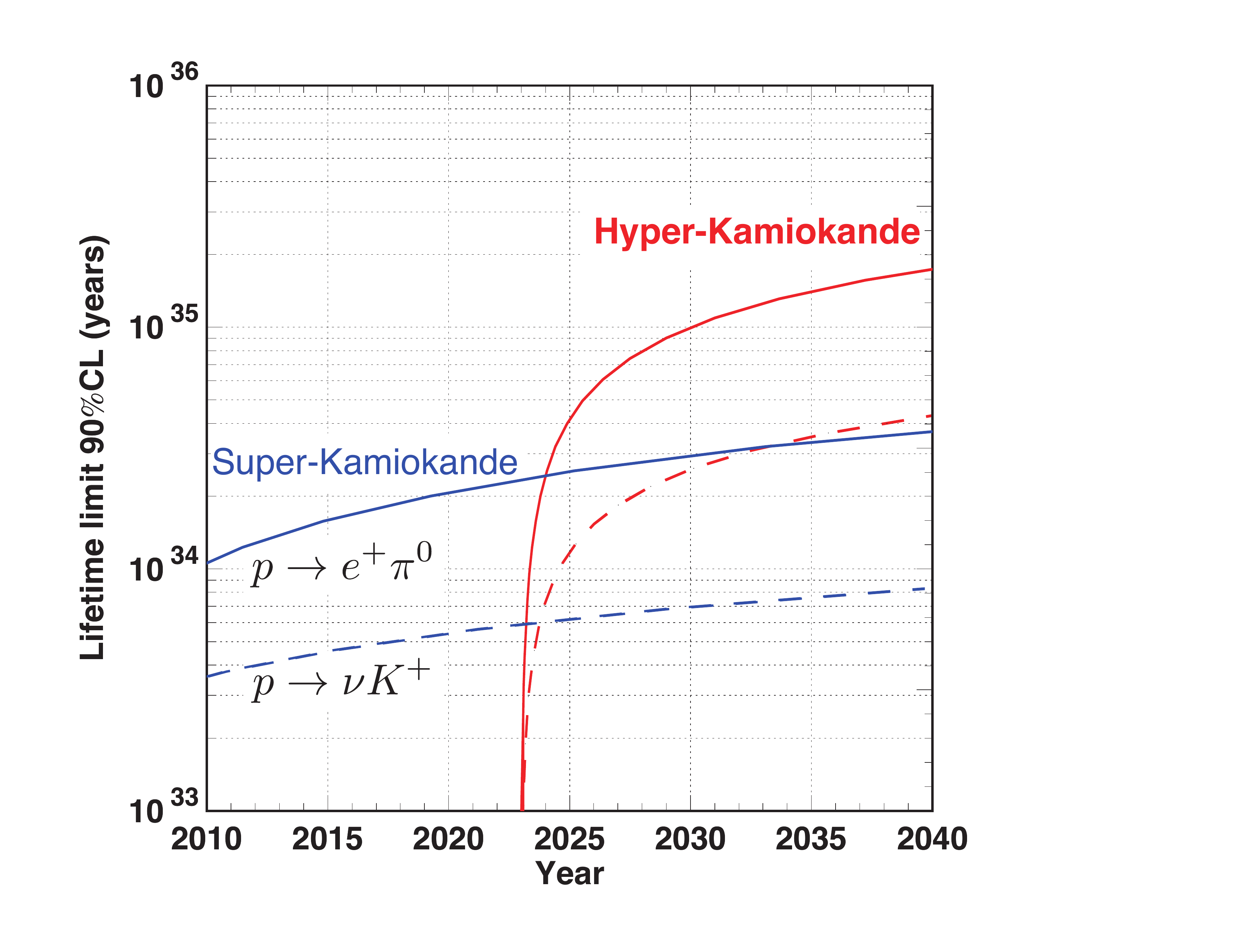}
  \end{center}
  \caption {Sensitivities of the Hyper-Kamiokande proton decay search as 
a function of detector exposure, at the 90\% C.L. The blue curves on the
left side shows the expected sensitivity for continued running of
Super-K; the red curves on the right show the sensitivity for Hyper-Kamiokande.
The upper solid curves are for $p \rightarrow e^+\pi^0$, for both experiments;
the lower solid curve are for $p \rightarrow \nu K^+$.}
\label{fig:hyperk_pdk_sens}                  
\end{figure}

The sensitivity of Hyper-K for nucleon decays has been studied based
on scaling of the the Super-Kamiokande analysis. The numbers for
signal efficiency and background rate from atmospheric neutrinos are
assumed to be identical to those used in the Super-K analyses reported
in Section~\ref{sec:superk-pdk}. Due to the very large mass and good
energy and Cherenkov ring reconstruction in the GeV range, water
Cherenkov is the best detector technology for signatures such as
$e^+\pi$ and $\mu^+\pi^0$. As was shown with Super-K analyses, the
efficiency and background rates are nearly unchanged if the photon
coverage is 20\% instead of 40\%. It is likely that the effective
photon coverage for Hyper-K will between those values. The planned
number of PMTs, 99000, corresponds to 20\% coverage; however recent
PMT manufacturing by Hamamatsu offers higher quantum efficiency than
previous generation PMTs such as those used in Super-K. It is safe to
assume an overall signal efficiency of 40\%, where the inefficiency is
dominated by nuclear interaction of the pion. For comparison, the
detection efficiency for decay of a free proton in H$_2$O is 87\%. The
background rate is well-established as discussed in the previous
section, and one can conservatively assume 2 events per 
Mton-years. Based on these numbers, the 90\% C.L. sensitivity of
Hyper-K for a 10-year exposure is greater than $10^{35}$ years, as
shown in Fig.~\ref{fig:hyperk_pdk_sens}.

Assuming the same analysis techniques employed by Super-Kamiokande,
one can estimate the sensitivity of Hyper-Kamiokande. Most sensitivity
comes from the two relatively background-free techniques:
$K^+ \rightarrow \pi^+\pi^0$ and $K \rightarrow \mu^+\nu$ with
nuclear-$\gamma$ tag. Based on the background rate in
Table~\ref{tab:sk-nukp-effbg}, a 10-year exposure would have an
expected background between 20 and 35 events. If the detected number
of events are equal to the background rate, the 90\% C.L. limit would
be roughly $3 \times 10^{34}$ years. This estimation assumes no
reoptimization of the analysis (tighter cuts) to accomodate the higher
background rate has been performed. Fig.~\ref{fig:hyperk_pdk_sens}
shows the 90\% CL sensitivity curve for the
$p \rightarrow \overline{\nu} K^{+}$ mode as a function of the
detector exposure.

\subsection{Liquid Argon TPC} 

The uniqueness of proton decay signatures in the LArTPC and the
potential for fully reconstructing the final state has long
been recognized as a strength for this technology. The LArTPC can
reconstruct all final state charged particles including an accurate
assessment of particle type, distinguishing muons from pions from
kaons from protons. Electromagnetic showers are readily measured with
a significant ability to distinguish those that originate from photons
from $\pi^0$ decay from those that originate from charged-current
electron neutrino interactions. Kiloton-per-kiloton, LAr TPC
technology will outperform water cherenkov in both detection
efficiency and atmospheric neutrino background rejection for most
nucleon decay modes, although intranuclear effects are smaller for
oxygen and non-existent for hydrogen.

Taking mass and cost into account, water Cherenkov technology is
optimum for the $p\to e^+\pi^0$ final state topology, where the signal
efficiency is roughly 40\% and the background rate is 2 events per
megaton-year. The estimate \cite{Bueno:2007um} for a LAr TPC is 45\%
efficiency and a background rate of 1 event per megaton year, not
enough of an improvement to overcome the penalty of lower mass.

\begin{figure}[htbp]
\centering
\includegraphics[width=0.6\textwidth]{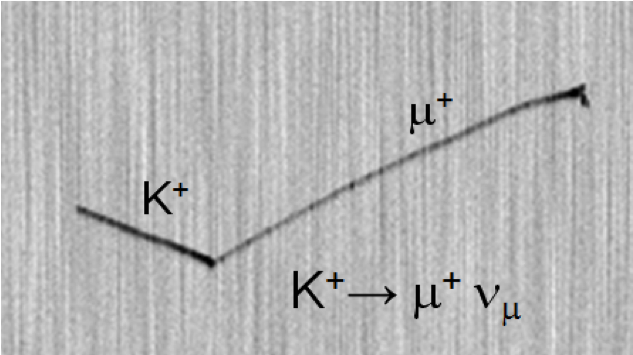}
\caption[Isolated Kaon observed during the ICARUS test run at Pavia]
  {Single event display for an isolated charged kaon in the ICARUS T600
   detector.  In this event, the kaon is observed as a heavily
   ionizing track that stops and decays to $\mu\nu$, producing a muon
   track that also stops and decays with a visible Michel electron.}
\label{fig:icaruskaon}
\end{figure}

On the other hand, for the $p \rightarrow \nu K^+$ channel, the
Super-K analysis yields a signal efficiency of roughly 19\% for a
background rate of 4 events per megaton year. This is the best mode
for a LArTPC, where the $K^+$ track is reconstructed and identified as
a charged kaon.  The charged kaon, with a momentum of 340 MeV/c
(neglecting nuclear effects), has a range of 14~cm in LAr, so
ionization energy loss measurements are expected to give high particle
identification efficiency. The decay products of the kaon are also
identified with high efficiency, for all kaon decay
branches. Fig.~\ref{fig:icaruskaon} shows an example event display of
an isolated charged kaon found in surface cosmic ray test run of the
ICARUS detector. The relative ionization of the kaon and muon are
readily evident, including the Bragg peak as the charged particle
comes to rest.

\begin{figure}[htbp]
\begin{center}
\includegraphics[width=0.8\textwidth]{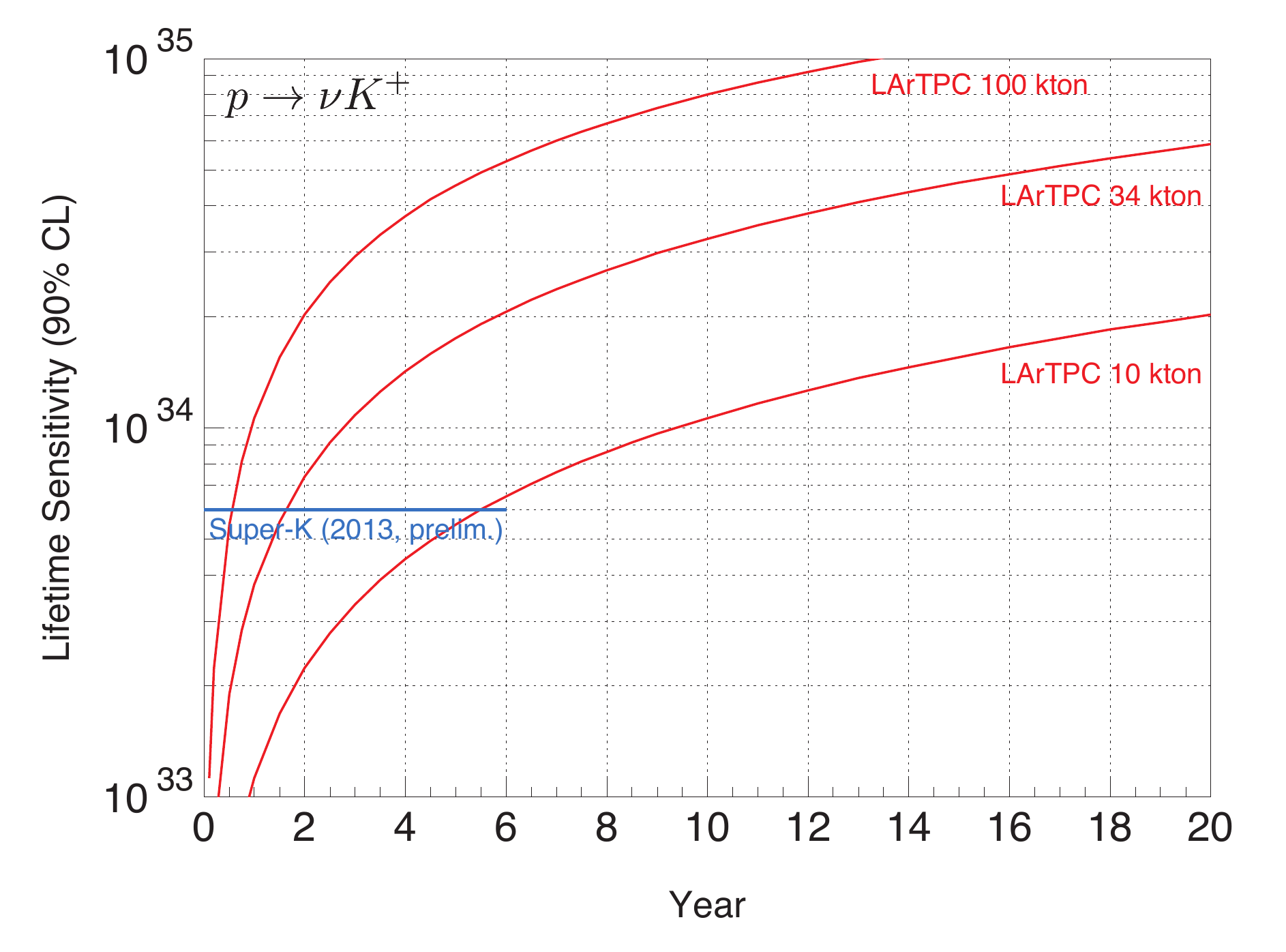}
\caption[Proton decay 90\% C.L. lifetime limit for $p \rightarrow K^+ \bar{\nu}$
  as a function of time]{Proton decay lifetime limit for
  $p \rightarrow K^+ \bar{\nu}$ as a function of time for various
  masses of a LArTPC, assumed to be sufficiently deep that a low
  background rate is achieved. For comparison, the 2013 preliminary
  result from Super-K for 11.6 live years is drawn. The projected
  limits are at 90\% C.L., calculated for a Poisson process including
  background assuming that the detected events equal the expected
  background.}
\label{fig:kdklimit}
\end{center}
\end{figure}

The total efficiency for the $\nu K^+$ mode is estimated to be as high
as 96.8\% with a background rate of 1 event per megaton year. Based on
these numbers and a ten year exposure, the 34 kton LBNE detector and
560 kton Hyper-Kamiokande have comparable sensitivity (at 90\% CL),
but the LArTPC would have an estimated background of 0.3 events
compared to tens of events for Hyper-K. Experimental searches for rare
events in the presence of significant backgrounds are notoriously more
problematic than background-free searches. A 10-year exposure of the
34-kton LBNE LArTPC detector would set a partial lifetime limit of
roughly $3 \times 10^{34}$ years, which is roughly the same as the
sensitivity of Hyper-K, despite 20 times fewer
proton-years. Figure~\ref{fig:kdklimit} shows the expected limit on
the proton lifetime as a function of time in LBNE for $p \rightarrow
\nu K^+$.

The LBNE LAr TPC has a chance to make up for lower detector mass when
compared to Hyper-Kamiokande for nucleon decay modes where a water
Cherenkov detector has relatively low efficiency or is susceptible to
higher background rates. Because the LAr TPC can reconstruct protons
that would otherwise be below Cherenkov threshold, it can reject many
CC and NC background topologies by vetoing on the presence of a recoil
proton. Because the LAr TPC has high spatial resolution, it does well
for event topologies with displaced vertices (such as $p \rightarrow
\mu^+ K^0$, a mode preferred in some SUSY GUTs over $\nu K^+$. For
modes with no electron in the final state, the same displaced vertex
performance we rely on for long-baseline neutrino oscillation allows
the rejection of charged current $\nu_e$ interactions. In general, the
above criteria favor nucleon decay modes with a kaon, charged or
neutral, in the final state. Conversely, the efficiency for decay
modes to a lepton plus light meson will be limited by intranuclear
reactions that are, if anything, worse than the case of $^{16}$O in a
water cherenkov detector.

An extensive survey of nucleon decay efficiency and background rates
has been published \cite{Bueno:2007um}. The table below lists selected
modes where a LAr TPC has a significant performance advantage (per
kiloton) over the water Cherenkov technique.

\begin{table}[htb]
\caption{Nucleon decay modes for which a LArTPC has an
advantage in signal efficiency and background rejection over a water
Cherenkov detector.}
\label{tab:lartpc-modes}
\begin{center}
\begin{tabular}{r|c|c|c|c} \hline\hline
 & WC & WC & LArTPC & LArTPC \\
decay mode & Efficiency & Bkg. Rate & Efficiency & Bkg. Rate \\ \hline
$p \rightarrow \nu K^+$ & 19\% & 4 /Mt/y & 97\% & 1 /Mt/y \\
$p \rightarrow \mu^+ K^0$ & 10\% & 8 /Mt/y & 47\% & $<$2 /Mt/y \\
$p \rightarrow \mu^- \pi^+ K^+$ &  &  & 97\% & $<$2 /Mt/y\\
$p \rightarrow e^- K^+$ & 10\% & 3 /Mt/y& 96\% & $<$2 /Mt/y\\
$p \rightarrow e^+ \pi^-$ & 19\% & 2 /Mt/y& 44\% & 0.8 /Mt/y\\
\hline
\hline
\end{tabular}
\end{center}
\end{table}

\subsection{Liquid Scintillator}

There remains a final experimental technique that may be used to
search for the favored proton decay mode $p \rightarrow \nu K^+$.  In
a water Cherenkov detector, the kaon is below Cherenkov threshold and
produces no Cherenkov light. In a scintillator detector, the kaon will
produce scintillation light with a yield of 100 or more photons per
MeV deposited. A fast scintillator and photomultipliers should be able
to resolve the sequential decay of the kaon ($tau \sim 12$ ns)
followed by the decay of the muon ($\tau \sim 2.2 \mu$s). The superior
energy resolution resulting from more collected photons will assist in
signal identification. Large (10$+$ kton scale) scintillator detectors
are envisioned for neutrino physics, astrophysics, and recently,
reactor neutrino oscillation (e.g Daya Bay 2). Except for this
particular decay mode, the proposed liquid scintillator detectors do
not have the mass or efficiency to compete with Hyper-Kamiokande or a
comparable size LArTPC for any other modes. However, the proposed
50-kton LENA detector would be competitive with Hyper-Kamiokande or a
34 kton LArTPC at LBNE for $p \rightarrow \nu K^+$.

LENA\cite{Wurm:2011zn} (Low Energy Neutrino Astronomy) is an
unsegmented liquid-scintillator detector of 50-kt target mass proposed
within the European LAGUNO-LBNO design study. While the emphasis of
the LENA physics program is on low-energy neutrinos and anti-neutrinos
(E$<$100 MeV), the search for proton decay into kaons and
antineutrinos was one of the first items considered to play an
integral part in the LENA concept, since the visibility of the kaon's
energy deposition in the scintillator substantially increases the
detection efficiency in comparison to water Cherenkov detectors.
Monte Carlo simulations show that analysis cuts can be defined which
retains a detection efficiency of $\sim 65$\% for the proton decay
signal. The search in LENA is expected to be background-free for about
10 years, allowing a lifetime limit of $\tau > 4 \times 10^{34}$ yrs
(90\% CL) if no event is observed.

Whereas LENA requires a substantial investment in a new cavern and
detector, it is possible to envision a liquid scintillator upgrade to
Super-Kamiokande. A white paper\cite{WbLSc} suggests a water-based
scintillator targeting a future upgrade for the 22.5 kton
Super-Kamiokande. A ten-year exposure could reach a 90\%
C.L. sensitivity on the partial lifetime of $p \rightarrow \nu K^+$ of
$2 \times 10^{34}$ years given favorable signal efficiency and
background rejection.

\section{Neutron-Antineutron Oscillations}
In this section and the next we turn to the $|\Delta {\cal B}| = 2$ process, neutron--antineutron oscillations.
The physics motivations are outlined here, followed by the experimental techniques focussing on the potential
improvement over past free neutron beam experiment by as much as four orders of magnitude in
transition probability will be addressed.

\subsection{Physics Motivation for $n - \bar n$ Searches}
\label{nnbar:sec:physics}

Historically, the idea that neutron may be its own antiparticle was first conjectured in 1937~\cite{Majorana:1937vz}.  With the development of particle physics since that time and specifically the acceptance of baryon number as a good symmetry to understand observed nuclear phenomena, it is now commonly accepted that the neutron is not a Majorana fermion. However a tiny Majorana component to its mass that violates baryon number remains an intriguing possibility.  The early history of other physics ideas related to $n$-$\bar{n}$ oscillations is briefly discussed in~\cite{Okun:2013voa}.
A more detailed exposure to neutron-antineutron oscillation can be found in Ref. \cite{whitepaper}.

There are many compelling reasons to think that fundamental particle interactions violate
baryon number. Arguably, the most powerful reason is that generating the
origin of the matter-anti-matter asymmetry in the universe requires that baryon
number must be violated~\cite{Sakharov:1967dj}.  Cosmological inflation, which is
strongly supported by astronomical data, implies that baryon number $\mathcal{B}$
was not conserved in the early universe~\cite{Dolgov:1991fr,Dolgov:1998ad}.  This
argument depends on the observed magnitude of the baryon asymmetry of the universe
but not on a mechanism of its generation.  Other reasons including grand unified theories,
has been addressed in the overview section.

Once we accept the possibility that baryon number is not a good symmetry of nature, there
are many questions that must be explored to decide the nature of physics associated with
$\mathcal{B}$-violation:  Is  $\mathcal{B}$
a global or local symmetry?  Does baryon number occur as a symmetry by itself or does it appear in combination with
lepton number, $\mathcal{L}$, i.e. $\mathcal{B}$ - $\mathcal{L}$, as the Standard Model
 would suggest? What is the scale of baryon number violation and the nature of the
associated physics that is responsible for it? For example, is this physics characterized
by a mass scale not too far above the TeV scale, so that it can be probed in experiments
already searching for new physics in colliders as well as low energy rare processes? Are
the details of the physics responsible for baryon-number violation such that they can
explain the origin of matter?

Proton decay searches probe baryon number violation due to physics at a grand
unified scale of $\sim 10^{15}-10^{16}$ GeV.  In contrast, the baryon-number violating
process of $n$-$\bar{n}$ oscillation, where a free neutron spontaneously
transmutes itself into an anti-neutron, has very different properties and
probes quite different physics.  The process violates baryon number by two
units and is caused by operators that have mass dimension nine so that it
probes new physics at mass scales $\sim$TeV and above. Therefore it can be probed by
experiments searching for new physics at this scale. It may also be deeply
connected to the possibility that neutrinos may be Majorana fermions, a natural
expectation.  A key question for experiments is whether there are theories that
predict $n$-$\bar{n}$ oscillations at a level that can be probed in
currently available facilities such as reactors or in contemplated ones such as
Project X, with intense neutron fluxes.  Equally important are the resulting
constraints on physics beyond the Standard Model if no signal appears after the
free-neutron oscillation time is improved by two orders of magnitude above the current
limit of 0.86$\times$10$^8$ s~\cite{BaldoCeolin:1994jz}.

%
\subsection{Some Background Concerning Baryon Number Violation}
\label{nnbar:subsec:theorybkgd}

Early on, it was observed that in a model with a left-right symmetric
electroweak group, $G_{LR} = {\rm SU}(2)_L \otimes {\rm SU}(2)_R \otimes {\rm
U}(1)_{B-L}$, baryon and lepton numbers in the combination $\mathcal{B}$ - $\mathcal{L}$ can be be
gauged in an anomaly-free manner. The resultant U(1)$_{B-L}$ can be combined
with color SU(3) in an SU(4) gauge group \cite{Pati:1974yy}, giving rise to the group
$G_{422} = {\rm SU}(4) \otimes {\rm SU}(2)_L \otimes {\rm SU}(2)_R$
\cite{Pati:1974yy,Mohapatra:1974hk,Mohapatra:1974gc}. A higher degree of unification involved models that embed either
the Standard Model gauge group $G_{SM} = {\rm SU}(3)_c \otimes {\rm SU}(2)_L
\otimes {\rm U}(1)_Y$ or $G_{422}$ in a simple group such as $SU(5)$ or $SO(10)$
~\cite{GG,GeorgiSO(10)}.  The motivations for grand unification theories
are well-known and include the unification of gauge interactions and their
couplings, the related explanation of the quantization of weak hypercharge and
electric charge, and the unification of quarks and leptons.
The unification of quarks and leptons in
grand unified theories generically leads to the decay of the proton and
the decay of neutrons that would otherwise be stably bound in nuclei. These
decays typically obey the selection rule $\Delta \mathcal{B} = -1$ and $\Delta \mathcal{L} = -1$.
However, the general possibility of a different kind of baryon-number violating
process, namely the $|\Delta \mathcal{B}|=2$ process of $n - \bar n$ oscillations, was
suggested~\cite{Kuzmin:1970vk} even before the advent of GUTs.  This was further
discussed and studied after the development of GUTs in~\cite{Glashow:1979sg,Mohapatra:1980qe} and
in a number of subsequent models~\cite{Kuo:1980ew,Chang:1980ey,Mohapatra:1980de,Cowsik:1980np,Rao:1982gt,
Misra:1982mg,Rao:1983sd,Huber:2001ug,Babu:2001qr,Nussinov:2001rb,Babu:2006xc,Dutta:2005af,Babu:2008rq,
Mohapatra:2009wp,Gu:2011ff,Babu:2013yca,Arnold:2012sd}.
Recently, a number of models have been constructed that predict $n - \bar n$ oscillations
at levels within reach of an improved search~\cite{Babu:2001qr,Nussinov:2001rb,Dutta:2005af,Babu:2013yca}.


\subsection{General Formalism for Analysis of $n- \bar n$ Oscillations}
\label{nnbar:subsec:formalism}

Since the neutron and antineutron have opposite magnetic moments, one must account for the magnetic splittings that may
be present between $n$ and $\bar{n}$ states in an oscillation experiment.  This motivates the following review
of the formalism for the two level ($n$,$\bar{n}$) system and $n - \bar n$ oscillations in an external magnetic
field~\cite{Mohapatra:1980qe}.

The $n$ and $\bar n$ interact with the external $\vec B$ field via their
magnetic dipole moments, ${\vec \mu}_{n,\bar n}$, where
$\mu_n = -\mu_{\bar n} = -1.9 \mu_N$ and $\mu_N = e/(2m_N) =
3.15 \times 10^{-14}$ MeV/Tesla.  Hence, the effective Hamiltonian matrix for the two-level $n - \bar n$ system takes the form
\begin{equation}
\cal{M}_{B}=\left(\begin{array}{cc}
m_n - {\vec \mu}_n \cdot {\vec B} - i\lambda/2 & \delta m \\
\delta m         & m_n + {\vec \mu}_n \cdot {\vec B} -  i\lambda/2
\end{array}\right),
\end{equation}
where $m_{n}$ is the mass of the neutron, $\delta m$ is the $\mathcal{B}$-violating potential coupling the $n$
and $\bar{n}$ states, and 1/$\lambda$ = $\tau_{n}$ = 880.1 $\rm{s}$ is the mean neutron lifetime.

The transition probability can be derived as
$P(n(t)=\bar n) = \sin^2(2\theta) \, \sin^2 [(\Delta E)t/2] \, e^{-\lambda t}$,
where $\Delta E \simeq 2 |\vec{\mu}_n\cdot\vec{B}|$ and $\tan(2\theta) = - \delta m/(\vec{\mu}_n\cdot\vec{B})$.
In a free propagation experiment, the quasi-free condition must hold, such that $|{\vec \mu}_n \cdot {\vec B}|t << 1$.
In this limit and also assuming that $t << \tau_{n}$, $P(n(t)=\bar n) \simeq [(\delta m) \, t]^2 = (t/\tau_{n- \bar n})^2$.

Then the number of $\bar n$'s produced by the $n - \bar n$ oscillations is given
essentially by $N_{\bar n}=P(n(t)=\bar n)N_n$, where $N_n = \phi T_{run}$, with
$\phi$ the integrated neutron flux and $T_{run}$ the running time.  The sensitivity of the
experiment depends in part on the product $t^2 \phi$, so, with adequate
magnetic shielding, one wants to maximize $t$, subject to the condition that
$|{\vec \mu}_n \cdot {\vec B}|t << 1$.


\subsubsection{$n - \bar n$ Oscillations in Matter}
\label{nnbar:subsubsec:matter}

To put the proposed free propagation $n$-$\bar{n}$ oscillation
experiment in perspective, it is appropriate to review limits that
have been achieved in the search for $n - \bar n$ oscillations in
nuclei. In 2002, the Soudan experiment reported a bound on the
intranuclear transition time of $\tau_m > 0.72 \times 10^{32} \ {\rm
  yr} \ (90 \% \ {\rm CL})$~\cite{Chung:2002fx} in $^{56}$Fe. Using
the relation $\tau_{n- \bar n} = \sqrt{ \tau_m/R} \ , $ where nuclear
structure calculations provide the suppresion factor $R \simeq
1.5\times10^{-23}$~s$^{-1}$, this is equivalent to $\tau_{n- \bar n}
\sim 1.3 \times 10^8$~s.  In 2011, the Super-Kamiokande experiment
reported a limit $\tau_m > 1.9 \times 10^{32} \ {\rm yr} \ (90 \%
\ {\rm CL})$~\cite{Abe:2011ky} in $^{16}$O, yielding $\tau_{n- \bar n}
\sim 3.5 \times 10^8$ s~ using the most recent value for the $R$
parameter for $^{16}$O \cite{Friedman:2008es}.  The envisioned free
neutron propagation experiment has the potential to improve
substantially on these limits. Achieving sensitivities of $\tau_{n-
  \bar n} \sim 10^9$ s to $10^{10}$ s would be roughly equivalent to
%
$\tau_m \simeq (1.6-3.1 \times 10^{33} \ {\rm yr})  ( \tau_{n- \bar n}/10^9 \ {\rm s}  )^2$.
%


\subsection{Operator Analysis and Estimate of Matrix Elements}
\label{nnbar:subsec:analysis}

At the quark level, the $n \to \bar n$ transition is
$(u d d) \to (u^c d^c d^c)$.  This is
mediated by six-quark operators ${\cal O}_i$, so the transition amplitude is
characterized by an effective mass scale $M_X$ and is expressed as
\begin{equation}
\delta m = \langle \bar n | H_{eff} | n \rangle = \frac{1}{M_X^5}
\sum_i c_i \langle \bar n |{\cal O}_i  | n \rangle .
\end{equation}
Hence,
$\delta m \sim \kappa \Lambda_{QCD}^{6}/M_X^{5}$,
where $\kappa$ is a generic $\kappa_i$ and $\Lambda_{QCD} \simeq 200$ MeV
arises from the matrix element $\langle \bar n | {\cal O}_i | n \rangle$.  For
$M_X \sim {\rm few} \ \times 10^5$ GeV, one has $\tau_{n- \bar n} \simeq 10^9 \
{\rm s}$.

The operators ${\cal O}_i$ must be color singlets and, for $M_X$
larger than the electroweak symmetry breaking scale, also ${\rm SU}(2)_L \times
{\rm U}(1)_Y$-singlets.  An analysis of these (operators) was carried out in~\cite{Rao:1982gt}
and the $\langle \bar n | {\cal O}_i | n \rangle$ matrix elements were
calculated in the MIT bag model.  Further results were obtained varying MIT bag
model parameters in~\cite{Rao:1983sd}.  These calculations involve integrals over
sixth-power polynomials of spherical Bessel functions from the quark
wavefunctions in the bag model.  From the arguments above,
it was found that
\begin{equation}
|\langle \bar n | {\cal O}_i | n \rangle | \sim O(10^{-4})
\ {\rm GeV}^6 \simeq (200 \ {\rm MeV})^6 \simeq \Lambda_{QCD}^6
\end{equation}
An exploratory effort has recently begun to calculate these matrix elements
using lattice gauge theory methods \cite{Buchoff:2012bm}.  Given that the mass scales probed
by these measurements go well beyond the TeV scale, the fundamental impact of a result (whether
or not oscillations are observed) and the availability of a variety of models predicting $n$-$\bar{n}$ at current sensitivity
levels ($\tau_{n- \bar n}\sim 10^{8}$ s), there is strong motivation to pursue a higher-sensitivity
$n - \bar n$ oscillation search experiment that can achieve a lower bound of $\tau_{n- \bar n} \sim
10^9 - 10^{10}$ s.

\section{NNbarX: An Experimental Search for $n- \bar n$ Oscillations at Project X}
\label{nnbar:sec:neutronpx}

Project X presents an opportunity to probe $n$-$\bar{n}$ transformation
with free neutrons with an unprecedented improvement in sensitivity~\cite{Kronfeld:2013uoa}.  Improvements would
be achieved by creating a unique facility, combining a high intensity
cold neutron source {\it dedicated} to particle physics experiments with
advanced neutron optics technology and detectors which build on
the demonstrated capability to detect antineutron
annihilation events with zero background.  Existing slow neutron sources at research reactors
and spallation sources possess neither the required space nor the degree of access
to the cold source needed to take full advantage of advanced neutron
optics technology which enables a greatly improved free $n$-$\bar{n}$ transformation
search experiment. Therefore, a dedicated source devoted exclusively
to fundamental neutron physics, such as would be available at Project
X, represents an exciting tool to explore not only $n$-$\bar{n}$ oscillations,
but also other Intensity Frontier questions accessible through slow neutrons.

\subsection{Previous Experimental Searches for $n- \bar n$ Oscillations}
\label{nnbar:subsec:previous}

As mentioned in Sec.~\ref{nnbar:subsubsec:matter}, the current best limit on $n$-$\bar{n}$ oscillations comes from the
Super-Kamiokande experiment, which determined an upper-bound on the
free neutron oscillation time of $\tau_{n-\bar{n}} >$ 3.5$\times10^{8}$ s
from $n$-$\bar{n}$ transformation in $^{16}$O nuclei~\cite{Friedman:2008es,Abe:2011ky}.
An important point for underground water Cherenkov measurements is that these experiments
are already limited in part by atmospheric
neutrino backgrounds.  Because only modest increments in detector mass
over Super-Kamiokande are feasible and the atmospheric neutrino backgrounds will scale
with the detector mass, dramatic improvements in the current limit
will be challenging for such experiments.

Experiments which utilize free neutrons to search for $n$-$\bar{n}$
oscillations have a number of remarkable features.  The basic idea for
these experiments is to prepare a beam of slow (below room temperature) neutrons which propagate
freely from the exit of a neutron guide to a distant annihilation target.  During
the time in which the neutron propagates freely, a $\mathcal{B}$-violating interaction can
produce oscillations from a pure ``$n$" state to one with an admixture of ``$n$" and ``${\bar n}$" amplitudes. Antineutron appearance is sought through
annihilation in a thin target, which generates several secondary
pions seen by a tracking detector situated around the target.  This signature
strongly suppresses backgrounds.  To observe this
signal, the ``quasi-free" condition must hold, in which the $n$ and $\bar{n}$
are effectively degenerate.  This creates a requirement for low pressures
(below roughly $10^{-5}$ Pa for Project X) and very small ambient magnetic fields
(between 1 and 10 nT for Project X) in order to prevent splittings between the
neutron and antineutron from damping the oscillations.  An improvement in sensitivity over the current free-neutron limit is available through the use of cutting-edge neutron optics, greatly increasing the neutron integrated flux and average transit time to the annihilation target.

\begin{figure}
    \centering \includegraphics[width=0.75\textwidth]
               {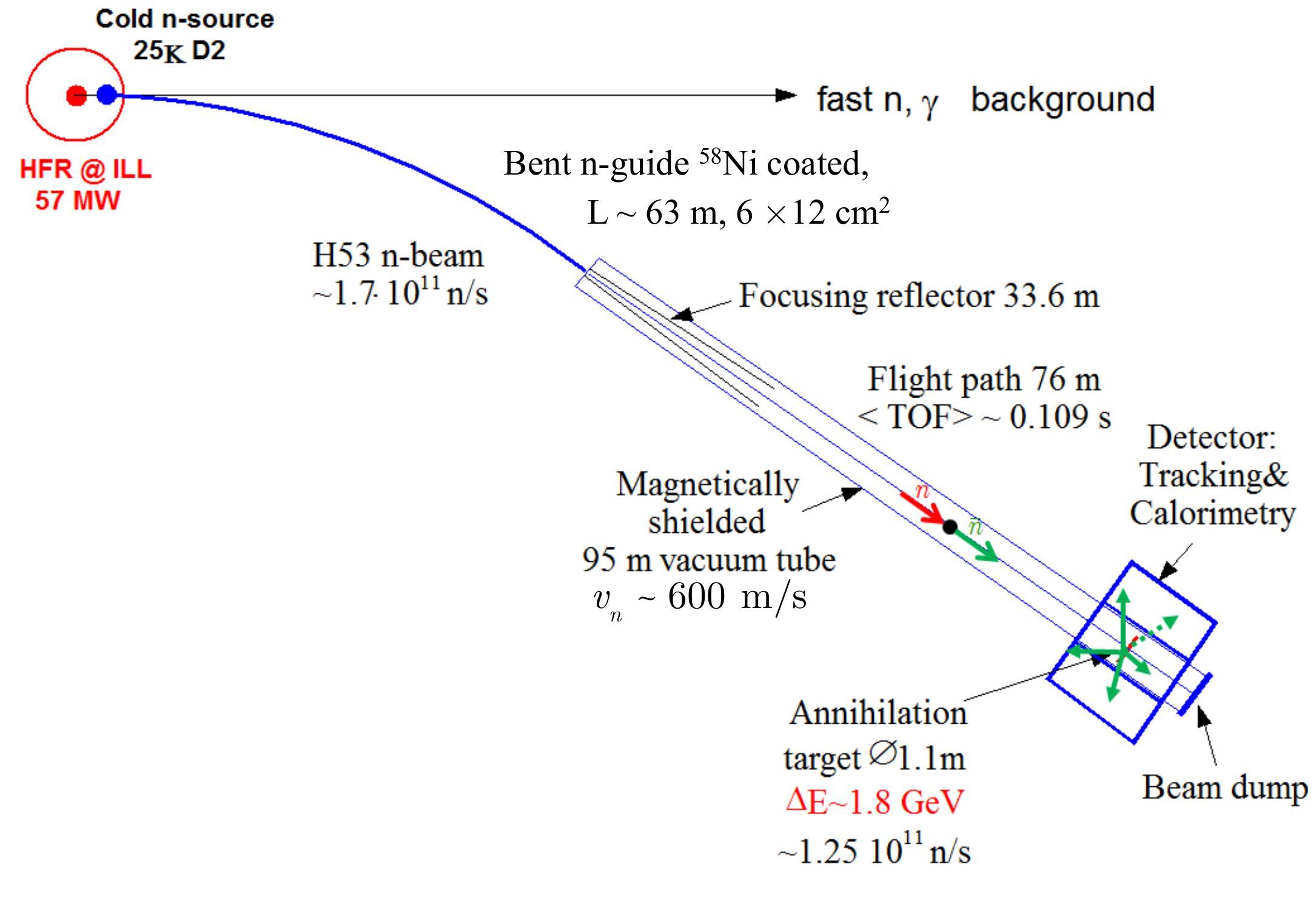}
    \caption{Configuration of the horizontal $n$-$\bar{n}$ search
    experiment at ILL/Grenoble~\cite{BaldoCeolin:1994jz}.}
    \label{ill:fig:logo}
\end{figure}

The current best limit for an experimental search for free $n$-$\bar{n}$
oscillations was performed at the ILL in
Grenoble in 1994~\cite{BaldoCeolin:1994jz} (see Fig.~\ref{ill:fig:logo}).
This experiment used a cold neutron beam from their 58 MW research reactor
with a neutron current of 1.25$\times$10$^{11} {\it n}/{\rm s}$
incident on the annihilation target and gave a limit of
$\tau_{n-\bar{n}} > 0.86\times10^{8}$ $\rm{s}$~\cite{BaldoCeolin:1994jz}.
The average velocity of the cold neutrons was $\sim$ 600 ${\rm m/s}$
and the average neutron observation time was $\sim$ 0.1 ${\rm s}$.
A vacuum of $P\simeq 2\times10^{-4}$ ${\rm Pa}$ maintained in the
neutron flight volume and a magnetic field of $|{\vec B}| < 10$ ${\rm nT}$
satisfied the quasi-free conditions for oscillations to occur.
Antineutron appearance was sought through annihilation with a $\sim$
130 ${\rm \mu m}$ thick carbon film target which generated at least
two tracks (one due to a charged particle) in the tracking detector with
a total energy above 850 MeV in the surrounding calorimeter. In one year of
operation the ILL experiment saw zero candidate events with zero
background~\cite{BaldoCeolin:1994jz}.

\subsection{Overview of the NNbarX Experiment}
\label{nnbar:subsec:overview}

A $n$-$\bar{n}$ oscillation search experiment at Project X (NNbarX) is
conceived of as a two-stage experiment. The neutron spallation
target/moderator/reflector system and the experimental apparatus need
to be designed together in order to optimize the sensitivity of the
experiment. The target system and the first-stage experiment can be
built and start operation during the commissioning of the first-stage
of Project X, which is based on a 1 GeV proton beam Linac
operating at 1 mA. The first-stage of NNbarX will be a horizontal
experiment with configuration similar to the ILL experiment~\cite{BaldoCeolin:1994jz} performed
in the 1990's, but employing
modernized technologies which include an optimized slow neutron
target/moderator/reflector system and an elliptical supermirror
neutron focusing reflector. Our very conservative baseline goal for a
first-stage experiment is a factor of 30 improvement of the
sensitivity (probability of appearance) for $n$-$\bar{n}$ oscillations
beyond the limits obtained in the ILL experiment~\cite{BaldoCeolin:1994jz}.
This level of sensitivity would also surpass the $n$-$\bar{n}$
oscillation limits obtained in the Super-Kamiokande, Soudan-II, and SNO intranuclear
searches~\cite{Abe:2011ky,Chung:2002fx,Bergevin:2010mb}.  In fact, although still in progress, our
optimization studies indicate that this horizontal geometry is capable
of improvements of a factor of 300 or more in 3 years of operation at
Project X.  A future, second stage of an NNbarX experiment can achieve higher
sensitivity by exploiting a vertical layout and a moderator/reflector
system which can make use of colder neutrons and ultracold neutrons
(UCN) for the $n$-$\bar{n}$ search.  This experimental arrangement
involves new technologies that will require a dedicated R$\&$D
campaign, but the sensitivity of NNbarX should improve by another
factor of $\sim$ 100 with this configuration, corresponding to limits
for the oscillation time parameter $\tau_{n-\bar{n}} > 10^{10}$
$\rm{s}$.  The increased sensitivity for $n$-$\bar{n}$ oscillations beyond
the ILL experiment~\cite{BaldoCeolin:1994jz} provide a strong motivation to search
for $n$-$\bar{n}$ oscillations as a part of Project X.

Intense beams of very low energy neutrons (meV) are available at
facilities optimized for condensed matter studies focused on neutron
scattering. These sources may be based on high flux reactors such as
the ILL or the High Flux Isotope Reactor (Oak Ridge) or on accelerator
based spallation sources such as SINQ
(Switzerland)~\cite{Blau:2009bb,Fischer:1997wf}, the SNS~\cite{Mason:2006tm},
or the JSNS in Japan~\cite{Maekawa:2010fk}.  Existing neutron
sources are designed and optimized to serve a large number of neutron
scattering instruments that each require beams with relatively small
cross sectional areas. A fully optimized neutron source
for an $n$-$\bar{n}$ oscillation experiment would require a beam having
a very large cross section and large solid angle. There are no such
beams at existing sources as these attributes would preclude them from providing the resolution
necessary for virtually all instruments suitable for materials
research. The creation of such a beam at an existing facility would
require major modifications to the source/moderator/shielding
configuration that would seriously impact the its efficacy for neutron
scattering.  The reason there has been no improvement
in the limit on free neutron $n$-$\bar{n}$ oscillations since the ILL
experiment is that no substantial improvement is possible
using existing sources.

The figure of merit for the sensitivity of a free
$n$-$\bar{n}$ search experiment is $N_{n}\cdot t^{2}$, where $N_{n}$
is the number of free neutrons observed and ${\it t}$ is the neutron
observation time (discussed in Sec.\ref{nnbar:subsec:formalism}).
The initial intensity of the neutron source was determined in the ILL
experiment by the brightness of the liquid deuterium cold neutron source
and the transmission of the curved neutron guide.  Although one expects the
sensitivity to improve as the average velocity of neutrons is reduced, it is
not practical to use very cold neutrons ($<$ 200 ${\rm m/s}$) with a horizontal
layout for the $n$-$\bar{n}$ search due to effects of Earth's gravity,
which will not allow free transport of very slow neutrons over significant
distances in the horizontal direction.

Modest improvements in the magnetic
field and vacuum levels reached for the ILL experiment would still
assure satisfaction of the quasi-free condition for the horizontal
experiment planned at Project X, but in our ongoing optimizations we
will investigate limits of $|{\vec B}|\leq 1$ ${\rm nT}$ in the whole free
flight volume and vacuum better than $P\sim 10^{-5}$ ${\rm Pa}$ in
anticipation of the more stringent requirements for a vertical
experiment. The costs of realizing these more stringent goals will be
considered in our optimization of the experimental design.

\begin{figure}
    \centering \includegraphics[width=0.99\textwidth]
               {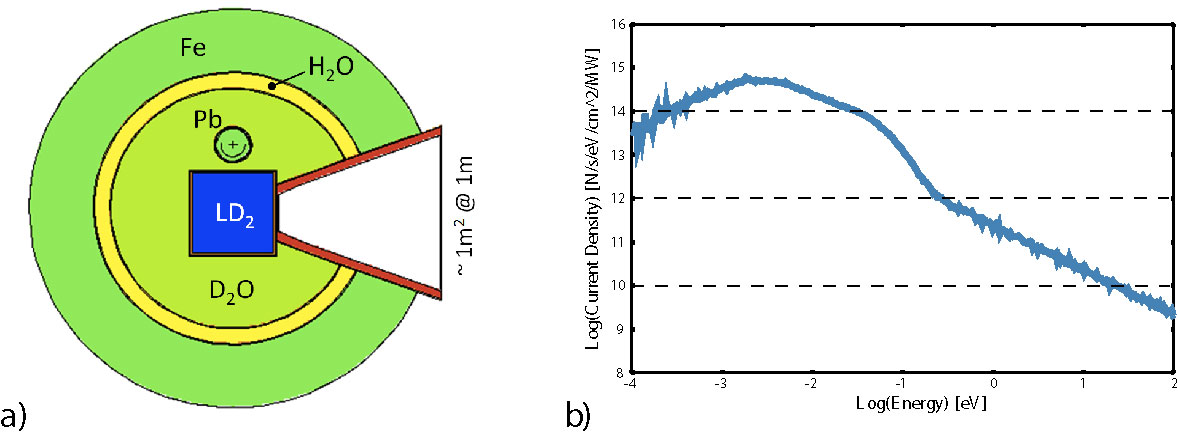}
    \caption{a) Depiction of the initial NNbarX baseline
cold neutron source geometry.  b) MCNP simulation of the cold neutron
spectrum entering the neutron optical system.}
    \label{nnbarx:fig:source}
\end{figure}

The Project X spallation target system will include a cooled
spallation target, reflectors and cold source cryogenics, remote
handling, nonconventional utilities, and shielding.  The delivery
point of any high-intensity beam is a target which presents
technically challenging issues for optimized engineering design, in
that optimal neutron performance must be balanced by effective
strategies for heat removal, radiation damage, remote handling of
radioactive target elements, shielding, and other aspects and
components of reliable safe operation.  The NNbarX baseline design incorporates
a spallation target core, which can be cooled by circulating water or
heavy water and will be coupled to a liquid deuterium cryogenic moderator
with optimized size and performance (see Fig. \ref{nnbarx:fig:source}).

\subsection{Increased Sensitivity of the NNbarX Experiment}
\label{nnbar:subsec:sensitivity}

A higher sensitivity in the NNbarX experiment compared to the previous
ILL experiment~\cite{BaldoCeolin:1994jz}, can be achieved by employing
various improvements in neutron optics and
moderation~\cite{Snow:2009zz}.  Conventional moderator designs can be
enhanced to increase the yield of cold neutrons through a number of
neutronics techniques such as a reentrant moderator
design~\cite{Mampe:1989fj}, use of
reflector/filters~\cite{Mocko:2013mm},  supermirror
reflectors~\cite{Swiss:2013sn}, and high-albedo materials such as
diamond nanoparticle
composites~\cite{Nesvizhevsky:2008rk,Lychagin:2009el,Lychagin:2008un}.
Although potentially of high positive impact for an $n$-$\bar{n}$
experiment, some of these techniques are not necessarily suitable for
multipurpose spallation sources serving a materials science user
community (where sharply defined neutron pulses in time may be
required, for example).

\begin{figure}
    \centering
    \includegraphics[width=0.99\textwidth]
                    {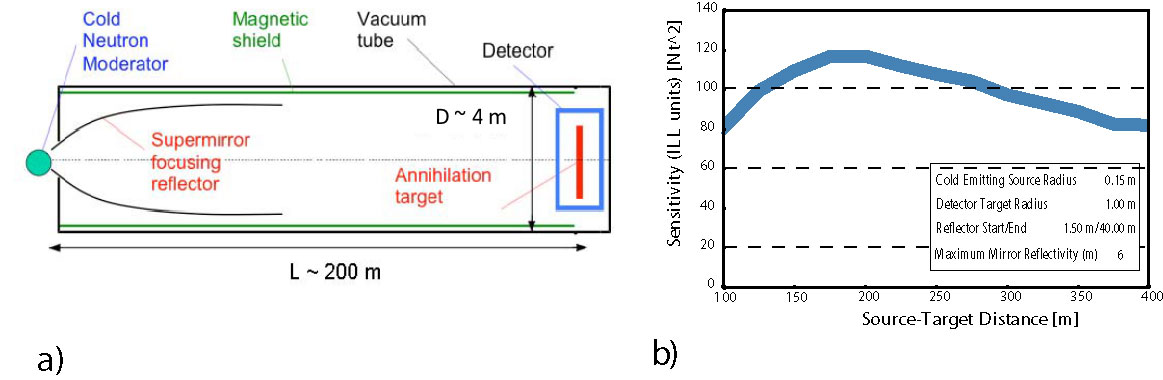}
    \caption{a) Schematic diagram of a candidate NNbarX geometry, depicting the relative
location of the cold neutron source, reflector, target, and annihilation detector.  b)  Calculation of the $n$-$\bar{n}$ oscillation
sensitivity for a geometry similar to that in panel (a), where all parameters are fixed except for the source-target distance.}
    \label{nnbarx:fig:sensitivity}
\end{figure}

Supermirrors based on multilayer coatings can greatly increase the
range of reflected transverse velocities relative to the nickel guides
used in the ILL experiment.  Supermirrors with $m = 4$, are now mass-produced and
supermirrors with up to $m = 7$, can be manufactured~\cite{Swiss:2013sn},
where ${\it m}$ is the near-unity reflection above nickel.  To enhance
the sensitivity of the $n$-$\bar{n}$ search, the
supermirrors can be arranged in the shape of a truncated focusing
ellipsoid~\cite{Kamyshkov:1995yk} (see Fig.~\ref{nnbarx:fig:sensitivity}a).
The focusing reflector with a large acceptance aperture will intercept neutrons
within a fixed solid angle and direct them by single reflection to the target.
The cold neutron source and annihilation target will be located in the focal planes of
the ellipsoid. The geometry of the reflector and the parameter ${\it
m}$ of the mirror material are chosen to maximize the sensitivity
$N_{n}\cdot t^{2}$ for a given source brightness and a given
moderator and annihilation target size.  Elliptical
concentrators of somewhat smaller scale have already been implemented
for a variety of cold neutron
experiments~\cite{Boni:2010pb}.  The plan to create a ${\it
dedicated}$ spallation neutron source for particle physics experiments
creates a unique opportunity to position the NNbarX neutron optical
system to accept a huge fraction of the neutron flux, resulting in
large gains in the number of neutrons directed to the annihilation
target.  Such a strategy makes use of a large fraction of
the available neutrons for a single beamline, so it would be incompatible
with a typical multi-user materials science facility.  Initial steps towards an optimized design have
been taken, with an NNbarX source design similar to the SINQ source
modeled and vetted vs. SINQ source performance (see
Fig. \ref{nnbarx:fig:source}), and a partially optimized elliptical
neutron optics system shown in Fig. \ref{nnbarx:fig:sensitivity}a.

MCNPX~\cite{Mcnpx:1981mc} simulation of the performance of the cold
source shown in Fig.~\ref{nnbarx:fig:source} produced a flux of
cold neutrons emitted from the face of cryogenic liquid  deuterium
moderator into forward hemisphere with the spectrum shown in
Fig.~\ref{nnbarx:fig:source}.  Only a fraction of the integrated flux is
accepted by the focusing reflector to contribute to the sensitivity at
the annihilation target.  Neutrons emitted from
the surface of neutron moderator were traced through  the detector
configuration shown in Fig.~\ref{nnbarx:fig:source} with gravity
taken into account and with focusing reflector  parameters that were
adjusted by a partial optimization procedure. The flux of cold
neutrons impinging on the annihilation detector target located at the
distance $L$ from the source was calculated after reflection (mostly
single) from the focusing mirror. The time of flight to the target
from the last reflection was also recorded in the simulation
procedure. Each traced neutron contributed to the total
sensitivity figure $N_n\cdot t^{2}$ that was finally normalized to the
initial neutron flux from the moderator. Sensitivity as function of
distance between neutron source and target is shown in
Fig.~\ref{nnbarx:fig:sensitivity}(b). The simulation has several parameters
that affect the sensitivity: emission area of the moderator, distance
between moderator and annihilation target, diameter of the
annihilation target, starting and ending distance for truncated
focusing mirror reflector, semi-major axis of the ellipsoid ($L$/2), and the
reflecting value ``$m$" of the mirror.  Sensitivity is a complicated
functional in the space of these parameters. A vital element of
our ongoing design work is to understand the projected cost for the
experiment as a function of these parameters.

A sensitivity in NNbarX in units of the ILL experiment larger than 100
per year of running seems feasible from these simulations.  Configurations of
parameters that would correspond to even larger sensitivities are
achievable, but for the baseline simulation shown in
Fig.~\ref{nnbarx:fig:sensitivity} we have chosen a set of parameters that
we believe will be reasonably achievable and economical after
inclusion of more engineering details than can be accommodated in our
simulations to date.  The optimal neutron optical configuration for an
$n$-$\bar{n}$ search is significantly different from anything that has
been built, so the impact on the sensitivity of cost
and engineering considerations is not simple to
predict at this early stage of the project. To demonstrate that the
key sensitivity parameters predicted by these
simulations do not dramatically depart from existing engineering
practice, we include Table~\ref{edm:tab:lqcd}, which shows the value of these same
parameters at existing MW-scale spallation neutron sources for the
source and optical parameters, and the ILL experiment for the
overall length $L$.

\begin{table}[ht]
\begin{threeparttable}
\centering
    	\caption{Comparison of parameters in NNbarX simulations with
    			existing practice.}
    	\label{edm:tab:lqcd}
    	\begin{tabular}{ccccc}
        \hline\hline Parameter & Units & Used in & Existing MW &
        References \\ & & Simulations & Facility Value & \\ \hline
        Source brightness & $n$/(s cm$^{2}$ sterad MW) &
        3.5$\times$10$^{12}$ & 4.5$\times$10$^{12}$ &
        ~\cite{Maekawa:2010fk} \\  ($E <$ 400 meV) &  &  &  & \\
        Moderator viewed area & cm$^{2}$ & 707 & 190 &
        ~\cite{Maekawa:2010fk} \\ Accepted solid
        angle\tnote{1} & sterad & 0.2 & 0.034 & ~\cite{Kai:2005tk} \\ Vacuum
        tube length & m & 200 & 100 & ~\cite{BaldoCeolin:1994jz} \\ $^{12}$C
        target diameter & m & 2.0 & 1.1 & ~\cite{BaldoCeolin:1994jz} \\
        \hline\hline
    \end{tabular}
    \begin{tablenotes}
    \item [1] Note that the solid angle quoted from JSNS
        is the total for a coupled parahydrogen moderator feeding 5
        neighboring beamlines (each of which would see a fifth of this
        value), whereas at NNbarX the one beam accepts the full solid
        angle.
    \end{tablenotes}
    \end{threeparttable}
\end{table}

\subsection{Requirements for an Annihilation Detector}
\label{nnbar:subsec:detector}

\begin{figure}[hbtp]
  \centering
    {\includegraphics[width=3.0in]
      {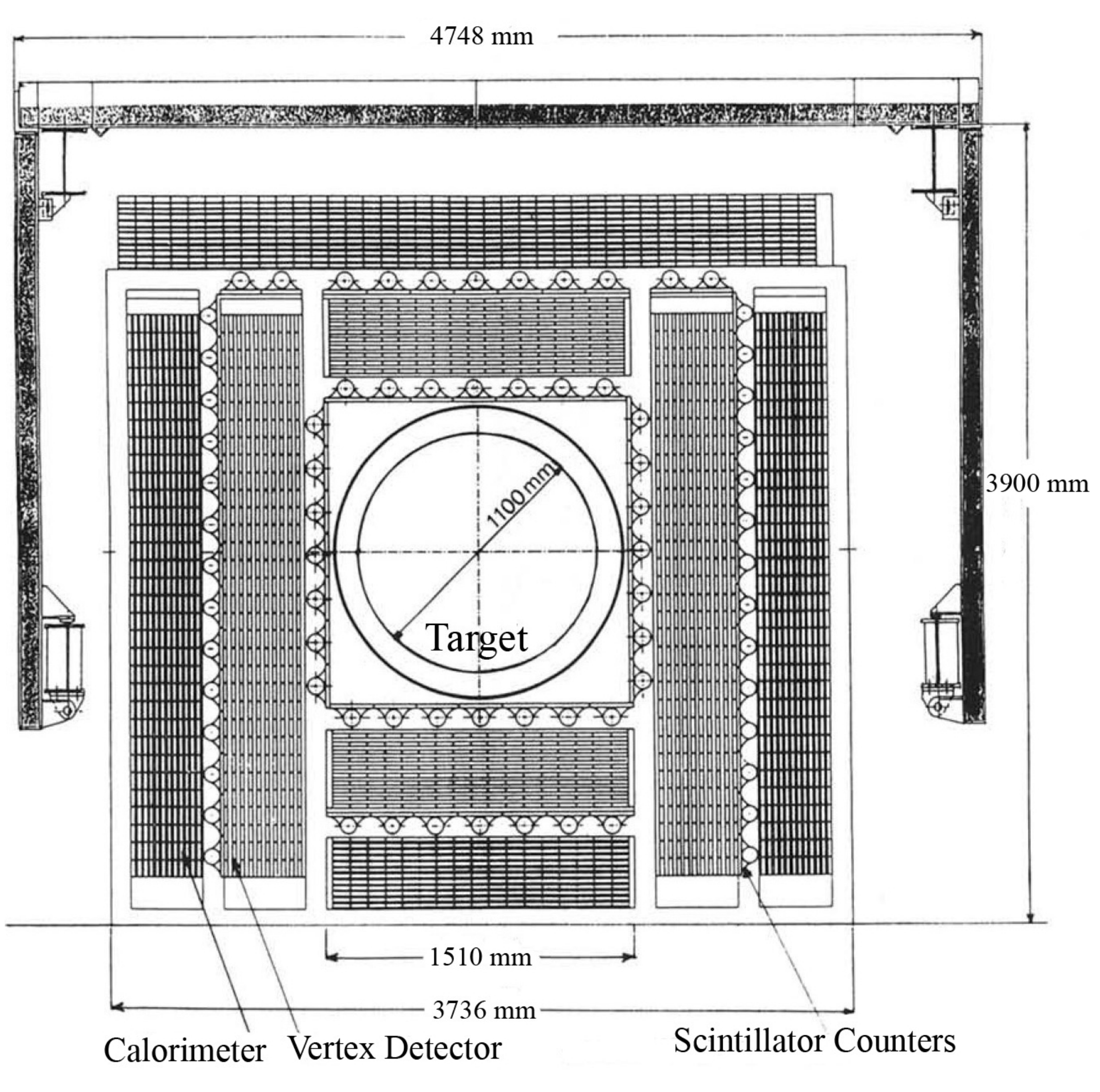}}
    {\includegraphics[width=3.0in]
      {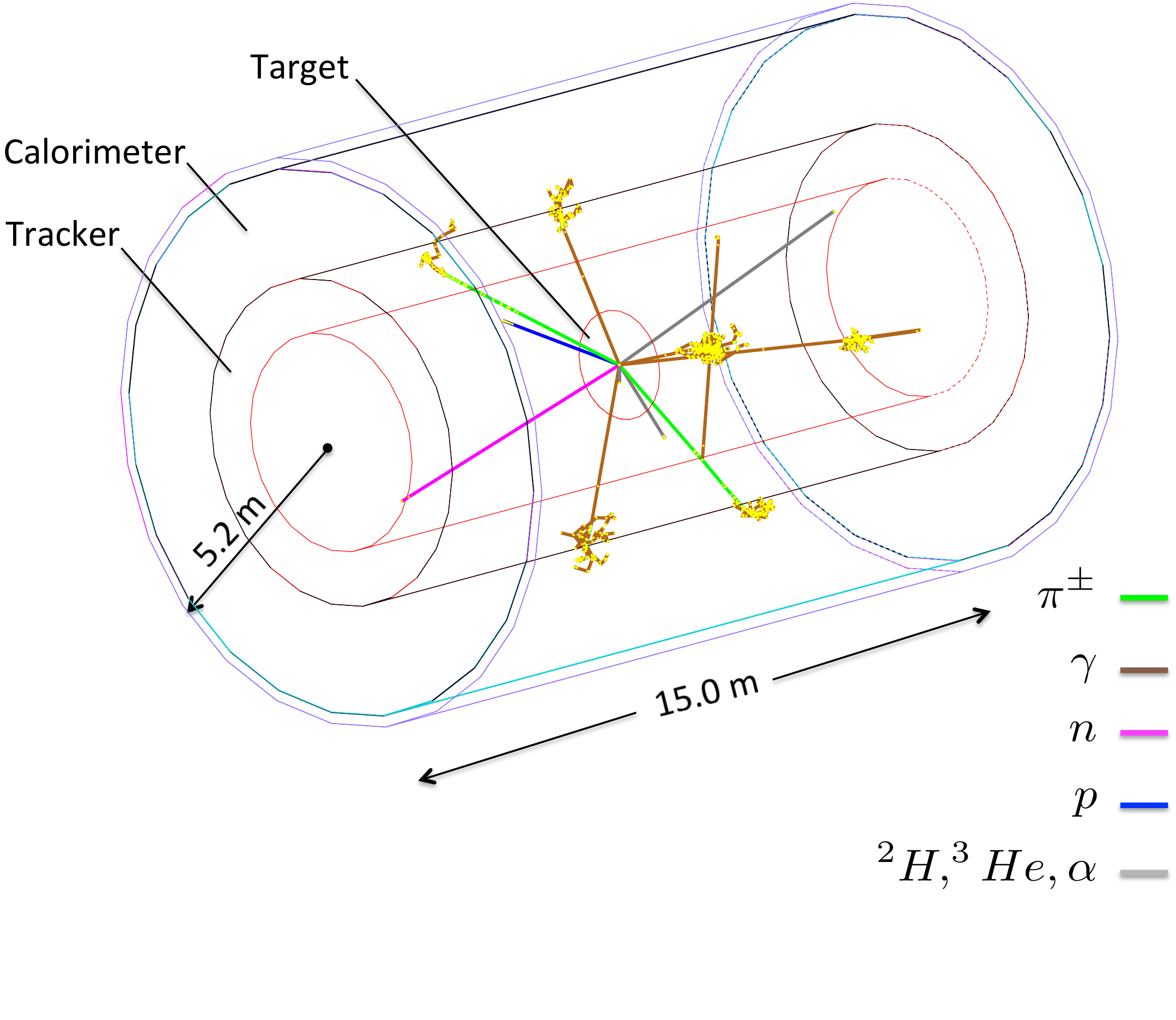}}
    \caption{a) Cross-sectional drawing of the ILL $n$-$\bar{n}$
    annihilation detector~\cite{BaldoCeolin:1994jz}.  b) Event display generated in our
    preliminary Geant4~\cite{Geant:2013ge} simulation for a $\pi^{+}\pi^{-}2\pi^{0}$
    annihilation event in a generalized NNbarX detector geometry.}
    \label{nnbarx:fig:detector}
\end{figure}

The target vacuum and magnetic field of 10$^{-5}$ Pa and $|{\vec B}| <$ 1 nT respectively is
achievable with standard vacuum technology and with an
incremental improvement on the ILL experiment through passive
shielding and straight-forward active field
compensation~\cite{BaldoCeolin:1994jz,Pxps:2012pp,Pxps:2012px}.  In the design of the
annihilation detector, our strategy is to develop
a state-of-the-art realization of the detector design used in the ILL
experiment~\cite{BaldoCeolin:1994jz} (see Fig. \ref{nnbarx:fig:detector}a).  The
spallation target geometry of NNbarX introduces a new
consideration in the annihilation detector design, because of the
possible presence of fast neutron and proton backgrounds.  These
backgrounds were effectively eliminated from the ILL
experiment, which produced fewer high energy particles in the reactor
source and eliminated the residual fast backgrounds using a curved
guide system to couple the cold source to the $n$-$\bar{n}$ guide.  For
NNbarX, we utilize a strategy of integrating our shielding scheme for
fast particles into the design of the source and beamline, and
optimize the choice of tracker detectors to differentiate between
charged and neutral tracks.  The residual fast backgrounds
at the detector are a strong function of the guide tube length,
detector threshold, and pulse structure for the proton beam.  In
particular, if needed, a slow chopping of the proton
beam (1 ms on, 1 ms off) will completely eliminate fast backgrounds
at the expense of the integrated flux of neutrons on target.

In general, the $n$-$\bar{n}$ detector doesn't require premium performance, but
due to relatively large size needs careful optimization of the
cost. In the current NNbarX baseline experiment, a uniform carbon disc in the
center of the detector vacuum region with a
thickness of $\sim$ 100 $\mu$m and diameter $\sim$ 2 m would serve as an annihilation
target.  Carbon is useful as an annihilation target due to the low capture cross
section for thermal neutrons $\sim$ 4 mb and high annihilation cross-section
$\sim$ 4 kb.  The fraction of hydrogen in the carbon film should be controlled
below $\sim$ 0.1$\%$ to reduce generation of capture $\gamma$'s.  The detector
should be built along a $\sim$ 4 m diameter vacuum region and cover a significant
solid angle in $\theta$-projection from $\sim$20$^{\circ}$ to 160$^{\circ}$ corresponding
to the solid angle coverage of $\sim$94$\%$.  The wall should be have a thickness
of $\sim$ 1.5 cm and be made of low-${\it Z}$ material (Al) to reduce multiple
scattering for tracking and provide a low (${\it n}$,$\gamma$) cross-section.
Additional lining of the inner surface of the vacuum region with $^{6}$LiF pads will
reduce the generation of $\gamma$'s by captured neutrons.  The detector vacuum
region is expected to be the source of $\sim$ 10$^{8}$ $\gamma$'s per second
originating from neutron capture.

A tracker system should extend radially from the outer surface of the
detector vacuum tube by $\sim$ 50 cm.  It should provide
rms $\leq$ 1 cm accuracy for annihilation vertex reconstruction to the
position of the target in the $\theta$-projection (compared to 4 cm in
ILL experiment). This is a very important resource for the control of
background suppression in the detector.  Reconstruction accuracy in the
$\phi$-projection can be a factor of 3 - 4 lower.  Relevant tracker
technologies can include straw tubes, proportional and drift
detectors.  A system similar to the ATLAS transition radiation tracker (TRT) is
currently under consideration for the tracking system.  The ATLAS TRT
has a measured barrel resolution of 118 $\mu$m and an end-cap resolution of 132 $\mu$m.  The
ATLAS TRT is capable of providing tracking for charged particles down
to a transverse momentum of $p_{T} =$ 0.25 GeV with an efficiency
of 93.6$\%$, but typically places a cut of $p_{T} >$ 1.00 GeV due to
combinatorics on the large number of tracks in collision events~\cite{Stahlman:2011js,
Boldyrev:2012ab,Vankooten:2013rv}.  The time of flight (TOF) systems should consist
of two layers of fast detectors (e.g. plastic scintillation slabs or tiles) before and after
the tracker.  With two layers separated by $\sim$50 cm - 60 cm, the TOF systems
should have timing accuracy sufficient to discriminate the annihilation-like tracks from the
cosmic ray background originating outside the detector volume.

The calorimeter will range out the annihilation products and should
provide trigger signal and energy measurements.  The average multiplicity of
pions in annihilation at rest equals 5, so an average pion can be
stopped in $\sim$20 cm of dense material (like lead or iron). For low
multiplicity (but small probability) annihilation modes, the amount of
material can be larger. The calorimeter configuration used in the ILL
experiment, with 12 layers of Al/Pb interspersed with gas detector
layers, might be a good approach for the
calorimeter design. Detailed performance for the measurement of total
energy of annihilation events and momentum balance in $\theta$- and
$\phi$-projections should be determined from simulations. An approach using
MINER$\nu$A-like wavelength shifting fibers coupled to scintillating bars is also being
considered~\cite{Mcfarland:2006km}.  The cosmic veto system (CVS) surrounding
the calorimeter should identify all cosmic ray
background.  Large area detectors similar to MINOS scintillator
supermodules~\cite{Michael:2008dm} might be a good approach to the
configuration of the CVS. Possible use of timing information should be
studied in connection with the TOF system. CVS information might not
be included in the trigger due to high cosmogenic rates, particularly
in the stage-one horizontal $n$-$\bar{n}$ configuration on the
surface, but should be recorded for all triggers in the off-line
analysis.

\subsection{NNbarX Simulation}
\label{nnbar:sec:simulation}

Developing a detector model through simulation that allows us to reach our goal of zero background
and optimum signal event detection efficiency is the primary goal of our simulation campaign,
which is currently underway.  We are using Geant 4.9.6~\cite{Geant:2013ge} to simulate the
passage of annihilation event products through the annihilation detector geometry with concurrent
remote development coordinated through GitHub~\cite{Github:2013gi}.  A detailed treatment of
$n$-$\bar{n}$ annihilation modes in $^{12}$C is currently under development.  According to
a Super-Kamiokande simulation study, 90$\%$ of the $n$-$\bar{n}$ annihilation modes in $^{16}$O
are purely pionic, while the remaining 10$\%$ are captured in the $\pi^{+}\pi^{-}\omega$
mode~\cite{Abe:2011ky}, which we expect to be similar to the physics of NNbarX.  The event
generator for $n$-$\bar{n}$ annihilation modes in $^{12}$C uses programs developed for the
IMB experiment and Kamiokande II collaborations~\cite{Jones:1983ij,Takita:1986zm} validated
in part by data from the LEAR experiment~\cite{Golubeva:1996eg}.  The branching ratios for
the $n$-$\bar{n}$ annihilation modes and fragmentation modes of the residual nucleus were
taken from Ref.~\cite{Abe:2011ky,Berger:1989gw,Fukuda:2002uc,Botvina:1990ke}.  The cross
sections for the $\pi$-residual nucleus interactions were based on extrapolation from measured
$\pi$-$^{12}$C and $\pi$-Al cross sections.  Excitation of the $\Delta$(1232) resonance was the
most important parameter in the nuclear propagation phase.  Nuclear interactions in the event
generator include $\pi$ and $\omega$ elastic scattering, $\pi$ charge exchange, $\pi$-production,
$\pi$-absorption, inelastic $\omega$-nucleon scattering to a $\pi$, and $\omega$ decays inside
the nucleus.  Fig. \ref{nnbarx:fig:detector}b shows an event display from our preliminary Geant4
simulation of a $\pi^{+}\pi^{-}2\pi^{0}$ annihilation event in a detector geometry with a
generalized tracker and calorimeter.


\subsection{The NNbarX Research and Development Program}
\label{nnbar:sec:randd}

In October of 2012, the Fermilab Physics Advisory Committee strongly
supported the physics of NNbarX and recommended that ``R$\&$D be
supported, when possible, for the design of the spallation target, and
for the overall optimization of the experiment, to bring it to the
level required for a proposal to be prepared.''   The NNbarX collaboration
has identified several areas where research and development may
substantially improve the physics reach of the experiment: target and
moderator design, neutron optics optimization and the annihilation
detector design.  At the core of this
activity is integrating models for the source, neutron optics and
detectors into a useful tool for evaluating overall sensitivity to
annihilation events and fast backgrounds, and developing a cost
scaling model.

There exist a number of improvements for the target and moderator, which have already
been established as effective and might be applied to our baseline
conventional source geometry.  For example, one can shift from a ${\it
cannelloni}$ target to a lead-bismuth eutectic (LBE)
target~\cite{Wohlmuther:2011mw}, utilize a reentrant moderator
design~\cite{Mampe:1989fj}, and possibly use
reflector/filters~\cite{Mocko:2013mm}, supermirror
reflectors~\cite{Swiss:2013sn}, and high-albedo materials such as
diamond nanoparticle
composites~\cite{Nesvizhevsky:2008rk,Lychagin:2009el,Lychagin:2008un}.
At present, the collaboration envisions a program to perform neutronic
simulations and possibly benchmark measurements on several of these
possibilities, with high-albedo reflectors as a priority.  Although
the basic performance of neutron optics is established, optimizing the selection of
supermirror technology for durability ($vs$ radiation damage) and cost
could have a very large impact on the ultimate reach of the experiment.

The collaboration is currently using the WNR facility
at LANSCE to determine the detection efficiency and timing properties
of a variety of detectors from 10 MeV to 800 MeV neutrons.  Detectors
under evaluation include proportional gas counters, straw tubes and
plastic scintillators.  Evaluating different
available detector options and modernizing the annihilation detector
should improve the background rejection capability and permit reliable
scaling to more stringent limits for $n$-$\bar{n}$ oscillations.  The
main technical challenges for NNBarX are to minimize the cost of
critical hardware elements, such as the large-area super-mirrors,
large-volume magnetic shielding, vacuum tube, shielding of the
high-acceptance front-end of the neutron transport tube, and
annihilation detector components.  These challenges will be addressed
in the R$\&$D phase for the NNBarX experiment.

\subsection{Summary}
\label{nnbar:sec:summary}

Assuming beam powers up to 1 MW on the spallation target and that 1
GeV protons are delivered from the Project X linac, the goal of NNbarX
will be to improve the sensitivity of an $n$-$\bar{n}$ search
($N_{n}\cdot t^{2}$) by at least a factor 30 (compared to the previous
limit set in ILL-based experiment~\cite{BaldoCeolin:1994jz}) with a
horizontal beam experiment; and by an additional factor of $\sim$ 100
at the second stage with the vertical layout. The R$\&$D phase of the
experiment, including development of the conceptual design of the cold
neutron spallation target, and conceptual design and optimization of
the performance of the first-stage of NNbarX is expected to take 2-3
years.  Preliminary results from this effort suggest that an
improvement over the ILL experiment by a factor of more than 100 may
be realized even in this horizontal mode, but more work is needed to
estimate the cost of improvements at this level.  The running time of
the first stage of NNbarX experiment is anticipated to be 3 years. The
second stage of NNbarX will be developed depending upon the
demonstration of technological principles and techniques of the first
stage.


\section{Conclusions}

While yet to be seen, proton decay is an indispensable tool for
probing Nature at truly high energies. It remains as the missing piece of  evidence
for grand unification. The dramatic meeting of the three gauge couplings at
a scale of about $2 \times 10^{16}$~GeV, which is found to occur in the
context of low energy supersymmetry, and the tiny neutrino masses as
observed in the neutrino oscillation experiments, lend  strong support
to the idea of supersymmetric grand unification.
Moreover, grand unified theories that are in accord with the observed
masses and mixings of all fermions, including neutrinos, typically predict
proton lifetimes within a factor of five to 10 of current Super-Kamiokande
limits.
This is why an improved search for proton decay is now most  pressing.
This can only be done with a large
detector built deep underground.
Such a detector, coupled to a long-baseline intense neutrino beam
(as would be available from Fermilab), can simultaneously
sensitively study neutrino oscillations so as to shed light on
neutrino mixing parameters, mass-ordering, and most importantly
$CP$ violation in the neutrino system. And it can help efficiently study
supernova neutrinos. In short, such a detector would have a unique
multi-purpose value with high discovery potential in all three areas.
Building such a large underground detector coupled
to a long-baseline neutrino beam in the US, in a timely fashion, would not
only probe a set of fundamental issues in physics, but would
enable the US to assume a leadership position by having a stellar facility that
would be an asset to the world as a whole.


Neutron--antineutron oscillations probe a different sector with baryon number
violation satisfying the selection rule $|\Delta {\cal B}| = 2$.  Discovery
of $n-\overline{n}$ oscillations in the next generation experiments would have
a profound impact on the physics beyond the Standard Model.  It may also suggest
new low scale mechanisms of generating baryon asymmetry of the Universe. The potential
to improve the oscillation probability by about four orders of magnitude provides
a golden opportunity for a potential landmark discovery in science.

\def\Discussion{\setlength{\parskip}{0.3cm}\setlength{\parindent}{0.0cm}
     \bigskip\bigskip      {\Large {\bf Discussion}} \bigskip}\def\speaker#1{{\bf #1:}\ }
\def\endDiscussion{}











\end{document}